\documentclass{article}

\usepackage[english]{babel}
\usepackage[separate-uncertainty]{siunitx}
\usepackage[letterpaper,top=2cm,bottom=2cm,left=1.5cm,right=1.5cm,marginparwidth=1.75cm]{geometry}

\usepackage{amsmath}
\usepackage{graphicx}
\usepackage[colorlinks=true, allcolors=blue]{hyperref}
\usepackage{multicol}
\usepackage{float}
\usepackage{doi}

\usepackage{subcaption}
\usepackage[numbers]{natbib}
\usepackage{ulem}

\usepackage[svgnames]{xcolor}
\usepackage{appendix}

\newcommand{\authormark}{\textsuperscript{,a)}}

\newcommand{\blackurl}[1]{%
    {\hypersetup{urlcolor=black}\url{#1}}%
}

\usepackage{eso-pic}

\newcommand{\firstpagefootnote}[2][0pt]{%
  \AddToShipoutPictureBG*{%
    \AtTextLowerLeft{%
      \raisebox{#1}{%
        \parbox[b]{\dimexpr(\textwidth-\columnsep)/2\relax}{%
          \footnotesize
          \rule{\dimexpr(\textwidth-\columnsep)/6\relax}{0.4pt}\\%
          #2%
        }%
      }%
    }%
  }%
}

\makeatletter
\def\bstctlcite{\@ifnextchar[{\@bstctlcite}{\@bstctlcite[@auxout]}}
\def\@bstctlcite[#1]#2{\@bsphack
  \@for\@citeb:=#2\do{%
    \edef\@citeb{\expandafter\@firstofone\@citeb}%
    \if@filesw\immediate\write\csname #1\endcsname{\string\citation{\@citeb}}\fi}%
  \@esphack}
\makeatother

\title{Implementation and verification of the avalanche source in a 3D full-f particle-in-cell model of relativistic electrons for studies of tokamak disruptions}
\author{F. Wouters$^1$\authormark, H. Bergstr\"om$^1$, M. Hoelzl$^{1,2}$ , G.T.A. Huijsmans$^3$, J. van Dijk$^4$ and the JOREK team$^{\text{b)}}$}
\begin{document}
%\bstctlcite{IEEEexample:BSTcontrol}
\maketitle

    %\vspace{0.5cm}
    
    \normalsize
    
    % List of the affiliations depending on the numbers used in the list of authors
    
    $^1$ Max Planck Institute for Plasma Physics, Boltzmannstraße 2, 85748 Garching, Germany

    $^2$ Department of Physics and Astronomy, Chalmers University of Technology, Göteborg, SE-41296, Sweden

    $^3$ CEA, IRFM, F-13108, Saint Paul-lez-Durance, France

    $^4$ Technische universiteit Eindhoven (TU/e), Groene Loper 3, 5612 AE Eindhoven, the Netherlands
    
    \vspace{0.6cm}

\begin{abstract}
Tokamak disruptions may lead to the acceleration of some electrons to relativistic energies.
These so-called runaway electrons (REs) can multiply exponentially via knock-on collisions with thermal electrons. As the resulting RE avalanche is exponentially sensitive to the pre-disruption plasma current, multi-MA RE beams may form in large future devices, risking severe localized wall damage.  
In this work, an energy and momentum conserving knock-on collision operator is implemented in the 3D nonlinear MHD code JOREK for the full-f relativistic hybrid fluid-kinetic model that describes the REs using the particle-in-cell (PiC) approach both for full-orbit and drift-kinetic markers, which will enable accurate modeling of the RE phase-space dynamics in realistic 3D electromagnetic fields. 
Such a self-consistent treatment of the RE avalanche and competing losses in the stochastic fields of MHD-active plasmas is required to further the understanding of RE transport and phase-space dynamics in self-consistent interaction with the 3D plasma evolution, which is needed for developing reliable predictions as well as reliable mitigation methods.
To make such novel high-fidelity simulations computationally viable, a resampling technique was also implemented to restrict the number of markers. The avalanche model is verified using analytical expressions from literature and applied to a JET-like termination scenario, demonstrating its applicability to realistic 3D MHD active scenarios. Future work on porting to accelerated high-performance computing systems will be needed to cross the long time scales involved, e.g., in periodic termination and re-avalanching that could occur in large devices like ITER.

\end{abstract}

\firstpagefootnote[\baselineskip]{%
  $^{\text{a)}}$Corresponding author: \blackurl{fiona.wouters@ipp.mpg.de}\\%
  $^{\text{b)}}$For a list of JOREK team members see the author list of [Nuclear Fusion 66, 116006 (2026). doi:10.1088/1741-4326/ae6790].%
}
\begin{multicols*}{2}
\raggedcolumns
\section{Introduction}
Disruptions, i.e., major magneto-hydrodynamic (MHD) instabilities that result in the loss of plasma confinement, are an unavoidable part of tokamak operation that can cause severe damage to a reactor-scale tokamak or may severely limit the lifetime of plasma facing components (PFCs)~\cite{Hender2007,Ratynskaia2026}. 
The characterization and mitigation of disruption loads is still a major challenge to overcome for future large-scale tokamaks~\cite{Lehnen2015}. 
Not only the high heat and electromagnetic loads associated with a disruption, but also the potential acceleration of electrons to relativistic velocities due to the disruption-induced electric field, have the potential to cause severe damage to the machine. 

The friction force of electrons in a plasma decreases with increasing velocity $v$ as $\sim \frac{1}{v^2}$~\cite{helander2005}. 
This implies that \columnbreak if the electric field is high enough, some of the fast electrons will be accelerated to relativistic velocities, which is an effect that was already predicted in 1924~\cite{Wilson1924, Wilson1925} and is known as the runaway phenomenon.
In particular, the electric field must be higher than the critical field given by the following expression~\cite{Connor1975}:
\begin{equation}
    E_c = \frac{e^3 n_e \ln(\Lambda)}{4 \pi \epsilon_0^2 m_e c^2}.
\end{equation}
Here $e$ is the elementary charge, $n_e$ the electron density, $\ln(\Lambda)$ the Coulomb logarithm, $\epsilon_0$ the vacuum permittivity, $m_e$ the electron mass and $c$ the speed of light. 
For a given parallel electric field $E_\parallel > E_c$, electrons moving parallel to the magnetic field with momenta above the critical momentum $p_c = m_e c/\sqrt{E_\parallel/E_c -1}$ will be in the runaway regime~\cite{Connor1975,Hoppe2021a}. 
Primary sources for runaway electrons (REs) are Dreicer generation~\cite{Dreicer1959}, hot tail generation~\cite{Chiu1998, Harvey_2000, Helander_2004}, tritium decay and Compton scattering~\cite{Martinsolis2017,Breizman2019} where the last two are of importance during the nuclear operation of a device. 
\end{multicols*}
\begin{multicols}{2}
Once there are a few of these REs present in the plasma, they can exponentially multiply by transferring their energy to the bulk electrons via knock-on collisions~\cite{Sokolov1979, Jayakumari1993, Rosenbluth1997}. 
This process, also known as the RE avalanche, is approximately exponentially sensitive to the pre-disruption plasma current, which may lead to MA carrying RE beams in a large current tokamak like ITER~\cite{Lehnen2015}.
In such cases the avalanche greatly increases the efficiency of the conversion of magnetic energy into kinetic energy stored in the runaway beam, which scales with $W_\text{RE} \sim I_\text{RE}^2$, during long terminations~\cite{Martin-Solis2014}.
Impact of such RE beams with the PFCs can not only lead to severe surface melting, but could also potentially penetrate deep enough into the material to damage cooling pipes and cause water leaks~\cite{Ratynskaia2026}. Substantial damage was already observed in present devices in spite of the much lower RE currents.
Simulations aimed at studying the generation and losses of these REs are therefore of paramount importance to assess risks and investigate mitigation methods. In particular the potential re-avalanching of a RE beam after it was lost in a burst of MHD mode activity (called termination) may substantially increase the risk beyond present devices, where re-avalanching is not efficient due to the much lower plasma current.

This need for predictive and interpretive modeling has led to the development of a large number of models with different capabilities, complexities and computational costs associated to them~\cite{Ratynskaia2026}. 
There are a number of codes that can treat the REs in either a fluid-like manner (e.g. GO~\cite{Smith2006, Eriksson2004, Vallhagen2020} and the 1.5-D transport code coupling ASTRA-STRAHL~\cite{Linder2020}), a bounce-averaged manner (e.g. CQL3D~\cite{Chiu1998}, LUKE~\cite{decker2005} and ARENA~\cite{Eriksson2003}), a full kinetic manner (e.g. NORSE~\cite{Stahl2017a},  CODE~\cite{Landreman2014}, KORC~\cite{Carbajal2017} and ASCOT\textbackslash ASCOT5~\cite{Sarkimaki2016, Izzo2022}) or both fluid-like and kinetic as is the case for DREAM~\cite{Hoppe2021}, M3D-C$^1$~\cite{M3DC1} and JOREK~\cite{Hoelzl2021}.
Most of these codes are lower dimensional and therefore don't resolve the full 3D MHD dynamics that are crucial to understand the full details of the disruption and RE transport, although the groundwork has been laid to couple KORC to the MHD code NIMROD~\cite{Lopez2025}. 
Exceptions to this are the MHD codes M3D-C$^1$ and JOREK, which both include a RE fluid model as well as kinetic extensions for REs. 

The kinetic extension for relativistic electrons in the 3D nonlinear MHD code JOREK~\cite{Hoelzl2021} treats the REs as either full-orbit or drift kinetic particles using a full-f particle-in-cell (PiC) approach, while the bulk plasma is described as a conducting fluid.
In recent work this kinetic extension has been coupled to the MHD simulation for both kinetic models~\cite{Bergstroem2025, Liu2025}, creating a hybrid fluid-kinetic framework capable of self-consistently evolving the REs and the MHD during MHD active scenarios such as disruptions.   
A number of effects relevant to REs have already been included in this model such as the hot tail source~\cite{Puel_2026}, small-angle collisions and the radiation reaction force~\cite{Sarkimaki2018, Sarkimaki2022}, and the effects of partial screening (based on Ref.~\cite{Hesslow2018}). The latter has not published explicitly for the kinetic model, however, the implementation for the fluid RE model can be found in Ref.~\cite{Bandaru2024}. 
In this work we implement an energy and momentum conserving knock-on collision operator for kinetic REs, which, to the authors' best knowledge, yields the first model capable of simulating the RE avalanche kinetically while self-consistently evolving 3D MHD active fields. 

The knock-on collision operator results in an exponential growth of REs. 
Consequently, care must be taken to make it computationally efficient in a PiC approach.
Therefore, the implemented model consists of two parts: the knock-on collisions and a resampling procedure. 
Knock-on Coulomb collisions have already been investigated with respect to ions in scenarios relevant for inertial confinement fusion in Monte Carlo methods~\cite{Turrel2015}, although a different method will be used in this work. 
The knock-on collisions here will be described using M\o{}ller scattering~\cite{Moller1931, Moller1932}, where energy and momentum conservation is used to determine all parameters involved. 
The collision operator is therefore energy and momentum conserving, allowing for a more accurate capturing of the growth-rate and pitch-angle distributions than the more commonly used Rosenbluth-Putvinski~\cite{Rosenbluth1997} and Chiu-Harvey~\cite{Chiu1998} source terms.
The physics covered are thus more similar to the knock-on collision operator derived from the Boltzmann collision operator in Ref.~\cite{Embreus2018} and the collisions used in Ref.~\cite{Beidler2024} than the previously mentioned source terms.
The collisions lead to an exponential increase in markers, which renders it computationally unfeasible for longer timescales, necessitating the implementation of a resampling algorithm. 
The resampling will be used to decrease the number of markers, while ensuring that the distribution function the markers represent is approximately retained. 

The rest of the paper is structured as follows. 
A short summary of the kinetic extension for relativistic particles in JOREK will be given in section~\ref{sec:model}. 
This will be followed by a description of the implementation of the avalanche source in section~\ref{sec:avalanchesource} consisting of collisions (subsection~\ref{sec:moller}) and resampling (subsection~\ref{sec:resampling}). 
In section~\ref{sec:application} it will be shown that the implemented source can be used to simulate the transition from RE losses to avalanche gain as the flux surfaces reform in a JET-like scenario. 
Finally, a summary and conclusion will be given in section~\ref{sec:conclusion} along with some outlook to future work.

\section{The model}
\label{sec:model}
The JOREK code allows for the REs to be treated kinetically in order to more accurately follow the particle trajectories and evolve the phase space distribution, and it is this hybrid kinetic-MHD model for which the avalanche source described here is implemented. 
The REs are modeled kinetically using a full-f particle-in-cell (PiC) method where markers are used, which each have a given weight $w$ indicating the number of physical particles they represent. 
It can be chosen to either use the full orbit (FO) or the guiding center (GC) pusher and the implementation of these pushers is described in Ref.~\cite{Sommariva2018} to which the interested reader is referred for the full details. 
The pushers will be treated here shortly for easy reference. 

In the FO model the REs are evolved using the following set of relativistic equations
\begin{align}
    \dot{\mathbf{x}} &= \frac{\mathbf{p}}{m_e \gamma}, \\
    \dot{\mathbf{p}} &= e \left(\mathbf{E} + \frac{\mathbf{p}}{m_e\gamma}\times \mathbf{B} \right),
\end{align}
where the Lorentz factor $\gamma = \sqrt{1 + \left(\frac{\mathbf{p}}{m_ec}\right)^2}$, with $\mathbf{x}$ and $\mathbf{p}$ the marker position and momentum vectors, and $\mathbf{E}$ and $\mathbf{B}$ the electric and magnetic field respectively. 
The timestep needs to be small of the order of $\Delta t_\text{push} = 0.01 \cdot T_\text{gyro}$ to fully resolve the gyro-orbit \cite{Sommariva2018}.  
The gyration period is given by $T_\text{gyro} = 2\pi \frac{m_e \gamma}{eB}$ and ranges typically between $10^{-11}-\SI{e-8}{\second}$ depending on particle energies and the magnetic field strength~\cite{Bergstroem2025}. 
As a non-conservative scheme might result in low solution accuracy, a volume preserving algorithm which is a symplectic algorithm developed in Refs.~\cite{zhang2015, wang2016, he2016} is used. The FO treatement for REs is computationally particularly expensive, but results in the most accurate description of RE transport and can produce accurate RE deposition patterns on realistic 3D walls including information of the RE phase space distribution~\cite{Bergstroem2024}, which allows for predictive wall damage assessments~\cite{Ratynskaia2026b}.

The GC model is a first order energy-like relativistic GC model described in Refs.~\cite{tao2007,cary2009}:
\begin{align}
    \begin{split}
    \dot{\mathbf{X}} &= \frac{1}{\mathbf{b} \cdot \mathbf{B}^*} \Biggl( -e \mathbf{E} \times \mathbf{b} - p_\parallel \frac{\partial \mathbf{b}}{\partial t} \times \mathbf{b} \\
    &\quad + \frac{m_e \mu \mathbf{b} \times \nabla B  + p_\parallel \mathbf{B}^*}{m_e \gamma_\text{GC}} \Biggr), \\
    \end{split} \\
    \dot{p}_\parallel &= \frac{\mathbf{B}^*}{\mathbf{b} \cdot \mathbf{B}^*}\cdot \left( -e\mathbf{E} - p_\parallel \frac{\partial \mathbf{b}}{\partial t} - \frac{\mu \nabla B}{\gamma_\text{GC}} \right), \\
    \gamma_\text{GC} &= \sqrt{1 + \left ( \frac{p_\parallel}{m_e c} \right)^2 + \frac{2 \mu B}{m_e c^2}},
\end{align}
where $\mathbf{X}$ is the position of the guiding center, $p_\parallel$ is the momentum parallel to the magnetic field, $\mu = \frac{p_\perp^2}{2 m_e B}$ is the magnetic moment, $\mathbf{b} = \frac{\mathbf{B}}{B}$ the direction of the magnetic field and $\mathbf{B}^* = p_\parallel \nabla \times \mathbf{b} + -e \mathbf{B}$ the effective magnetic field. 
These equations are solved using the fifth order Cash-Karp Runge-Kutta scheme described in Ref.~\cite{cash1990}.
If needed, an adaptive time stepping allows for the mitigation of the lack of symplecticity in this scheme. The GC treatment of REs is computationally much more efficient than the FO model, still captures transport much more accurately than RE fluid models, and allows for detailed study of the effect of kinetic RE orbits on MHD modes~\cite{Liu2025}.

Both models include additional effects that can either be included or neglected such as: the radiation reaction force, small-angle collisions and partial screening (additional details for the extensions were shown on the previous page).
A recently added RE source is the hot tail source~\cite{Puel_2026}, which means that, with the inclusion of the avalanche source as described in this work, the kinetic RE model of JOREK includes the most important sources for non-nuclear operation.

\section{The avalanche source}
\label{sec:avalanchesource}
This section will treat the implemented avalanche source in two parts: first the knock-on collisions that make up the avalanche source are discussed in subsection~\ref{sec:moller}, followed by the resampling procedure that counteracts the exponential growth of markers in subsection~\ref{sec:resampling}.  

\subsection{Knock-on collisions}
\label{sec:moller} 
M\o{}ller scattering~\cite{Moller1931, Moller1932} is the relativistic description of electron-electron scattering and will therefore form the basis of the implementation of the knock-on collisions.
The thermal electrons will be assumed to be approximately stationary, which holds when the RE velocities are much larger than the thermal velocity~\cite{Rosenbluth1997, Chiu1998}. 
The differential M\o{}ller scattering cross-section for a collision with a stationary electron is given by~\cite{Breizman2019, Berestetskii1982}
\begin{multline}
\label{eq:diffMollerscatt}
     \frac{d \sigma}{d \gamma} = \frac{2 \pi r_e^2}{\gamma_0^2 - 1}\Biggl[ \left(\frac{\gamma_0}{\gamma -1} \right)^2 + \left(\frac{\gamma_0}{\gamma_0 - \gamma}\right)^2 +1  \\
      - \frac{2\gamma_0 -1}{\gamma_0 - 1} \left(\frac{1}{\gamma -1} + \frac{1}{\gamma_0 - \gamma}\right)\Biggr],
\end{multline}
where $r_e$ is the classical electron radius, $\gamma_0$ the Lorentz factor of the incoming RE and $\gamma$ the Lorentz factor of the electron after the collision.

The full integral of this differential cross-section will always diverge due to the inclusion of `collisions' in which no energy was transferred. 
Therefore, a cut-off momentum $\left(\gamma_\text{min} = \sqrt{1 + (\frac{p_\text{min}}{m_e c})^2}\right)$ has to be introduced, which will also be used to distinguish between small-angle and knock-on collisions. 
Due to energy conservation, an outgoing electron can then only have energies in the range $\gamma_\text{min} \leq \gamma \leq \gamma_0 +1 -\gamma_\text{min}$, giving the lower and upper bounds of the integral.
This allows for determining the total cross-section, which is then given by

\begin{multline}
\label{eq:totcross}
    \sigma = \frac{4 \pi r_e^2}{\gamma_0^2 -1}\Biggl[\gamma_0^2\frac{\gamma_0 - 2\gamma_\text{min}+1}{(\gamma_\text{min}-1)(\gamma_0 - \gamma_\text{min})} + \frac{1}{2}(\gamma_0 +1) \\
     - \gamma_\text{min} - \frac{2 \gamma_0 -1}{\gamma_0 -1} \ln\left(\frac{\gamma_0 - \gamma_\text{min}}{\gamma_\text{min}-1}\right)\Biggr].
\end{multline}

It will be assumed that the energy associated to the critical momentum $p_c$ is far larger than the ionization energy, such that both free and bound electrons are seen as targets for the knock-on collision.
A method that more carefully accounts for the binding energy of the electrons using a more accurate ionization cross-section for argon atoms (as derived in Ref.~\cite{Bretagne_1986}) has been proposed, but the effect of this correction tends to be small for the largest part of the RE generation during tokamak disruptions~\cite{McDevitt_2019}. 
As our main focus is on modeling disruption scenarios, this correction is left for future work. 

The probability of a RE with velocity $v_0$ to collide with a thermal background electron within a collision timestep $\Delta t$ is
\begin{equation}
\label{eq:prob}
    P \approx \frac{1}{2} \sigma n_e^\text{tot} v_0 \Delta t = \frac{1}{2} \sigma n_e^\text{tot} \frac{\sqrt{\gamma_0^2-1}}{\gamma_0} c \Delta t 
\end{equation}
which holds when $\frac{1}{2} \sigma n_e^\text{tot} v_0 \Delta t \ll 1$. 
Here $n_e^\text{tot} = n_e^\text{free} +n_e^\text{bound}$ is the total electron density.
The factor $\frac{1}{2}$ in Equation~\eqref{eq:prob} has to be added because the total cross-section describes the scattering of 2 electrons, while only 1 of these will be a new RE.
As each marker represents a large number of REs, the probability $P$ determines the fraction of the REs represented by the marker that collided. 
After the collision process, each colliding marker will split into three. The first one representing the fraction of REs that were not affected by the collision, the second one representing the newly formed REs by the collision, and the third one representing the fraction of incoming REs that lost energy in the collision process.
Assuming the weight of the initial marker to be $w_0$, it is known that $P w_0$ REs collided. This sets the weights of the second and third markers as $P w_0$, while the first marker describing the fraction of REs that didn't collide is set to $(1-P)w_0.$
Adding two new markers (as compared to only adding one new marker to account for the newly generated REs) allows us to capture the effect of the large angle collisions on the already existing RE population and conserve energy and momentum in the collisions, at the cost of a rapid increase in the number of markers.
Although the two electrons are indistinguishable after the knock-on collision, the RE with the lowest energy is often referred to as the secondary RE.

As the probability may not exceed 1, there is an upper limit on $\Delta t$ given by
\begin{equation}
\label{eq:deltat}
    \Delta t < \frac{2}{\sigma n_e^\text{tot} \frac{\sqrt{\gamma_0^2-1}}{\gamma_0} c} = \Delta t_\text{limit}.
\end{equation}
This means that the time taken between collisions is dependent on the density of the thermal bulk electrons and the energy distribution of the REs. 
Even though the time between collisions is constrained as given in Equation~\eqref{eq:deltat}, in general $\Delta t \gg \Delta t_\text{push}$ so that $\Delta t = n \Delta t_\text{push}$ where $n$ is an integer number. 

\begin{figure}[H]
    \centering
    \includegraphics[width=0.7\linewidth]{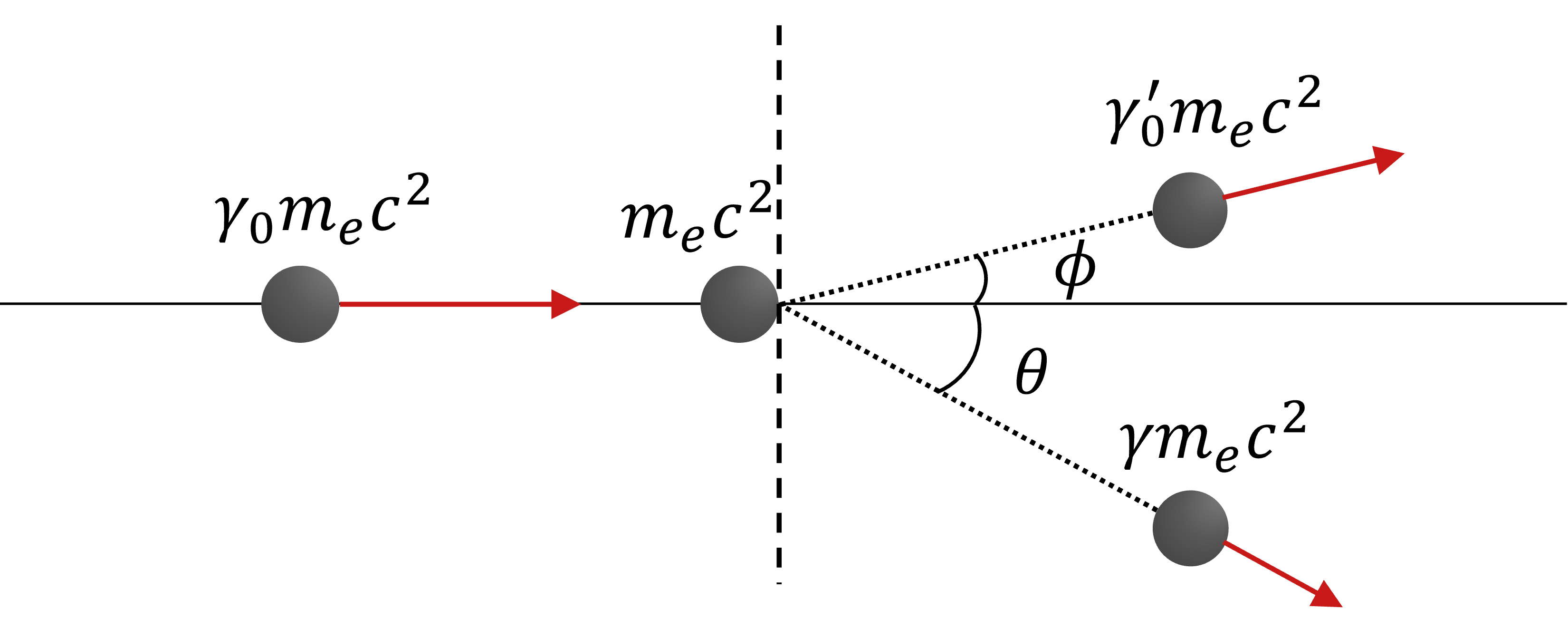}
    \caption{Illustration of the collision between a relativistic electron with energy $\gamma_0 m_e c^2$ and a stationary electron with rest energy $m_e c^2$. After the collision one electron will have an energy of $\gamma_0' m_e c^2$ and have a scattering angle $\phi$ while the other will have an energy of $\gamma m_e c^2$ and have a scattering angle $\theta$.}
    \label{fig:coll2D}
\end{figure}

The energy of one of the outgoing electrons $\mathcal{E} = \gamma m_e c^2$ is chosen proportional to the probability density $p(\gamma) = \frac{1}{\sigma} \frac{d \sigma}{d \gamma}$ using rejection sampling.
Energy conservation then sets the Lorentz factor of the other electron as $\gamma_0' = \gamma_0 + 1 - \gamma$. 
Once the energies of the outgoing electrons are known, momentum conservation can be used to describe the scattering angles in the 2D plane of the collision.
The angle $\theta$ that the momentum vector of the electron with $\mathcal{E}=\gamma m_e c^2$ makes with the initial momentum vector is given by
\begin{equation}
\label{eq:cospitch}
    \cos(\theta) = \sqrt{\frac{\gamma - 1}{\gamma + 1}} \sqrt{\frac{\gamma_0 +1}{\gamma_0 - 1}}, 
\end{equation}
while the angle $\phi$ of the other electron (with $\mathcal{E}_0'=\gamma_0' m_e c^2$) is
\begin{equation}
    \sin(\phi) = \sqrt{\frac{\gamma^2 -1}{ ( \gamma_0 + 1 - \gamma)^2 -1}} \sin(\theta).
\end{equation}
This fully describes the collision in the 2D plane as illustrated in figure~\ref{fig:coll2D}.
The 2D plane of the collision is set using the direction of the initial momentum vector and a randomly generated orthogonal vector.
As this 2D plane is chosen randomly with respect to the initial momentum vector, this can lead to multiple different pitch-angles for markers with respect to the magnetic field, even if the collisions are identical in every other aspect.
When the GC model is used instead of the FO model, the markers are first converted to FO markers and converted back to GC markers after the collision using an iterative approach (described in more detail in subsection~\ref{sec:resampling}). 

\subsubsection{Growth-rate}
For a mono-energetic population of REs with Lorentz factor $\gamma_0$ the increase in number of REs at time $t + \Delta t$ is given by
\begin{equation}
    N_\text{RE} (t + \Delta t) - N_\text{RE}(t) = \frac{1}{2} \sigma n_e^\text{tot} \frac{\sqrt{\gamma_0^2-1}}{\gamma_0} c \Delta t N_\text{RE}(t).  
\end{equation}
As the new REs are generated at the same position as the original population, the volume they inhabit can be considered constant, which means that the growth-rate in the limit of $\Delta t \rightarrow 0$ becomes
\begin{equation}
\label{gamma}
    \Gamma \equiv \frac{1}{n_\text{RE}} \frac{d n_\text{RE}}{d t} \approx \frac{1}{n_\text{RE}} \frac{\Delta n_\text{RE}}{\Delta t} = \frac{n_e^\text{tot} \sigma c}{2} \frac{\sqrt{\gamma_0^2-1}}{\gamma_0}.
\end{equation}
To first order this equals the theoretical growth-rate for a mono-energetic population as seen in Ref.~\cite{Aleynikov2015}, provided that the same cut-off momentum $\gamma_\text{min}$ is used and $\Delta t$ is chosen sufficiently small. 

\begin{figure}[H]
    \vspace{-0.7cm}
    \centering
    \includegraphics[width=1.1\linewidth]{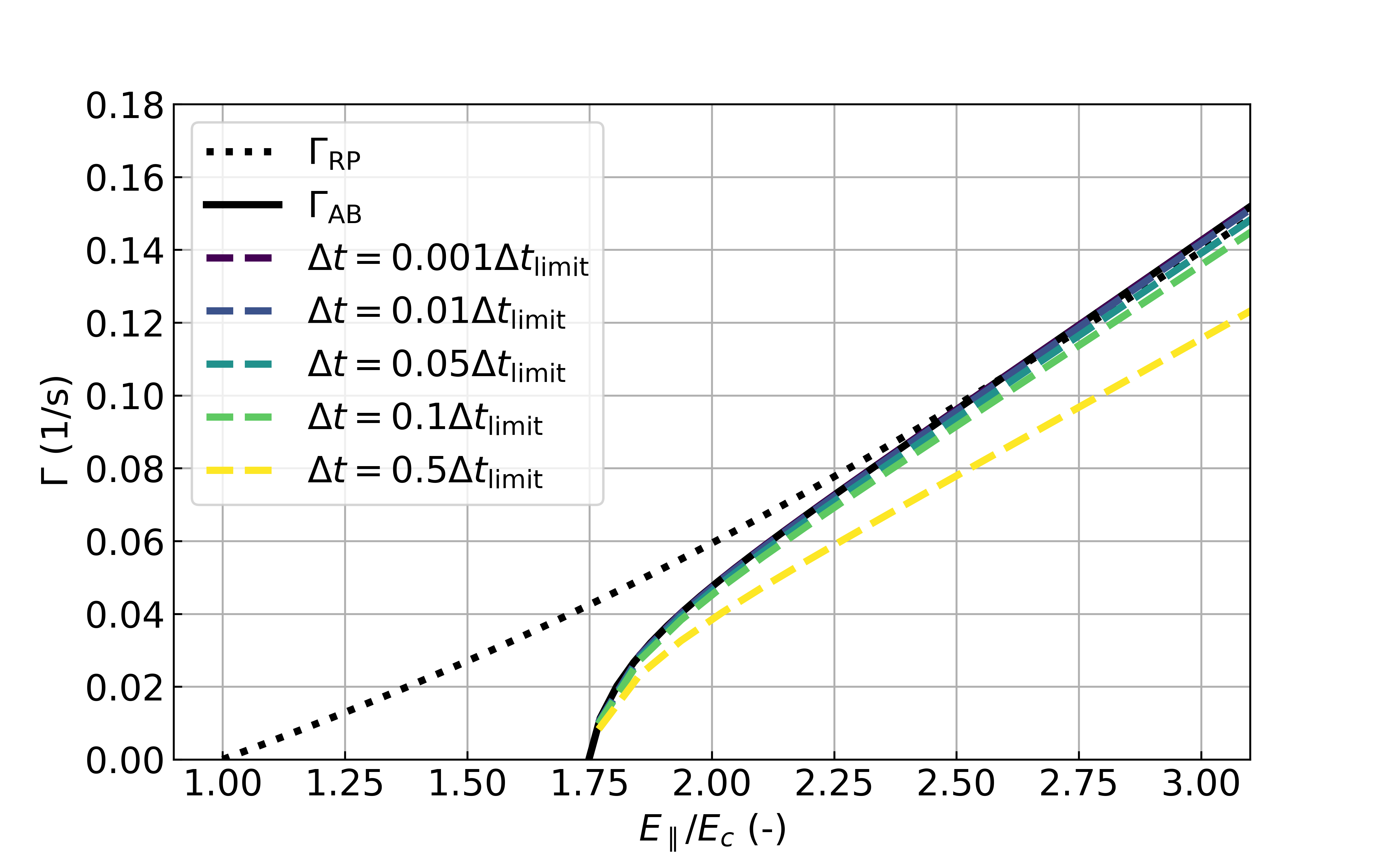}
    \caption[width=\linewidth]{The growth-rate resulting for different values of $\Delta t$ is shown to converge to the theoretical growth-rate from Ref.~\cite{Aleynikov2015} $\Gamma_\text{AB}$ when $\Delta t \leq 0.01 \Delta t_\text{limit}$. The Rosenbluth-Putvinski growth rate $\Gamma_\text{RP}$ is also shown for comparison.}
    \label{fig:growth_rate}
\end{figure}

\begin{figure*}[ht!]
     \centering
     \begin{subfigure}[b]{0.35\textwidth}
         \centering
         \includegraphics[height=0.85\textwidth]{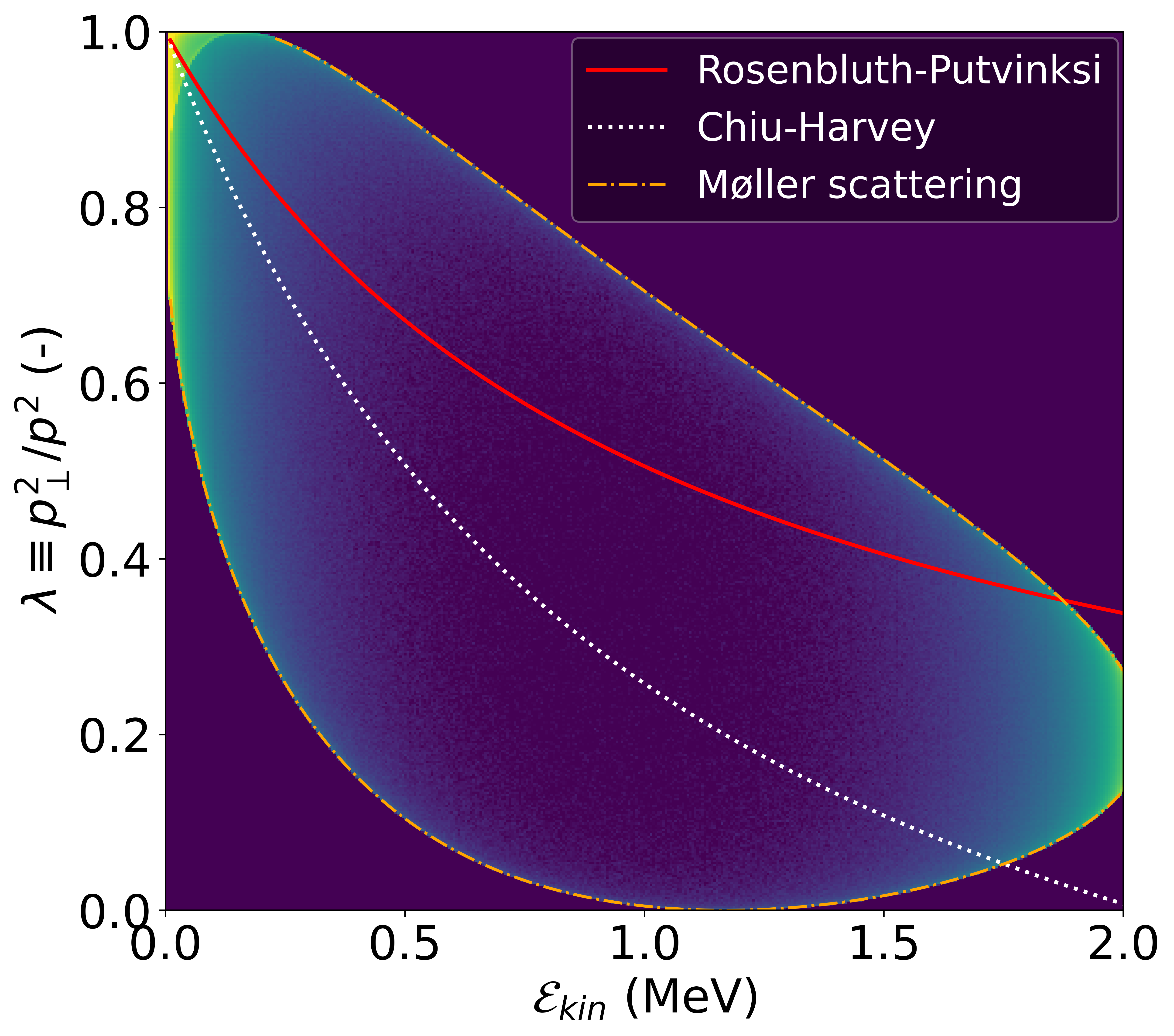}
         \caption{Knock-on collisions}
         \label{subfig:mysource}
     \end{subfigure}
     \begin{subfigure}[b]{0.35\textwidth}
         \centering
         \includegraphics[height=0.85\textwidth]{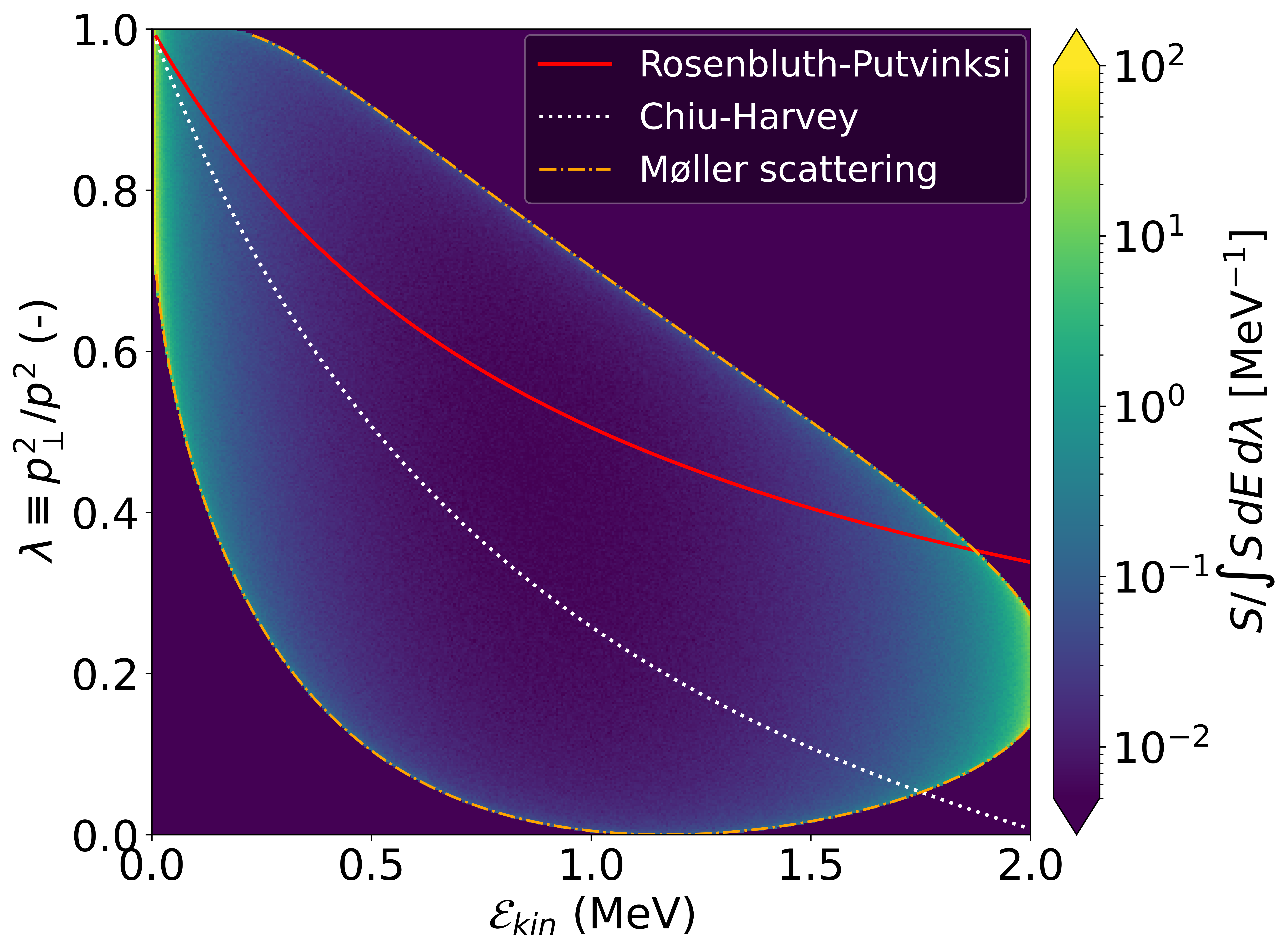}
         \caption{Full non-simplified source}
         \label{subfig:fullsource}
     \end{subfigure}
        \caption{Two normalized RE avalanche source terms are color-coded using a logarithmic scale colormap taking inspiration from figure 4 in Ref.~\cite{Aleynikov??}. The source term corresponding to primary electrons with $\gamma_0 = 5$ and $\lambda_0 \equiv \frac{p_{\perp,0}^2}{p_0^2} = 0.2$ resulting from the implementation shown here, is depicted in (a). The full non-simplified source operator (Equation (8) from Ref.~\cite{Aleynikov??}) is shown in (b). The red and white dotted lines in (a) and (b) represent the lines along which the Rosenbluth-Putvinski~\cite{Rosenbluth1997} and Chiu-Harvey~\cite{Chiu1998} sources generate REs in momentum space, respectively. Note, that the source term shows both the lower and higher energy outgoing electrons from the collision, as the incoming electron and the collision target are indistinguishable after the collision. The orange dashed line represents the bounds set by energy and momentum conservation for the two M\o{}ller scattered REs. The difference in the top left corner of (a) is due to the binary collision generating secondary electrons that move in the anti-parallel direction. These are removed from the simulation at a later point when their momentum drops below $p_c$ due to deceleration by the electric field.}
        \label{fig:momentum_space}
\end{figure*}

To illustrate the error that is introduced when $\Delta t$ is chosen too large, the growth rate is matched to the theoretical growth rate using the same $\gamma_\text{min}$ and $\gamma_0$ as was used in Ref.~\cite{Aleynikov2015}. 
For this test, an initial purely mono-energetic population of REs with energy $\gamma_0 m_e c^2$ is considered and the resulting instantaneous growth-rate is compared to the growth-rate from Aleynikov and Breizman $\Gamma_\text{AB}$ (Equation~(11) from Ref.~\cite{Aleynikov2015}). 
Figure~\ref{fig:growth_rate} shows that the growth-rate is well matched for $\Delta t \leq 0.01 \Delta t_\text{limit}$, where $\Delta t_\text{limit}$ is defined by Equation~\eqref{eq:deltat} and is in the order of 6 to 40 seconds for $n_e^\text{tot} =$ \SI{1e19}{\meter^{-3}} in this case.
These values for $\Delta t_\text{limit}$ are not representative for disruption cases in large machines such as ITER, where $\frac{E_\parallel}{E_c} \gg 1$ in general. 

\subsubsection{Generation in phase space}
The model presented here is based both on energy and momentum conservation and therefore differs from the commonly used Rosenbluth-Putvinski source~\cite{Rosenbluth1997}, which assumes the incoming RE to move parallel to the magnetic field with infinite energy. 
An improvement upon this source has been made by Chiu and Harvey~\cite{Chiu1998} by including the finite energy of the incoming RE while keeping the assumption that they move parallel to the magnetic field. 
The difference between these sources and the source implemented here can most easily be seen when the source is considered in momentum space. 
The pitch of the secondary electron generated by the Rosenbluth-Putvinski source is 
\begin{equation}
    \xi = \cos(\theta) = \sqrt{\frac{\gamma-1}{\gamma+1}}.
\end{equation}
This means that, in momentum space, the Rosenbluth-Putvinski source will always generate the secondary electrons along a line, as illustrated in figure~\ref{fig:momentum_space}. 
When the initial RE is parallel to the magnetic field but the energy is finite as is the case for the Chiu-Harvey source, the pitch angle is the same as Equation~\eqref{eq:cospitch} except that $\theta$ now represents the angle with respect to the magnetic field instead of the momentum of the initial RE. 
The Chiu-Harvey source would therefore generate along a line in momentum space (see figure~\ref{fig:momentum_space}) if the initial distribution is mono-energetic, or in a region determined by the energy range of the REs. 
The model implemented here takes the pitch $\xi_0$ of the original RE into account, however, such that REs are generated in a larger bounded region if $|\xi_0| \neq 1.0$ even if the original population is mono-energetic, as shown for $\gamma_0 = 5$ and $\lambda_0 \equiv \frac{p_{\perp,0}^2}{p_0^2} = 0.2$ in figure~\ref{subfig:mysource}.  
This matches with the full non-simplified source term as given in Refs.~\cite{Aleynikov??, Boozer2015} as shown in figure~\ref{subfig:fullsource}, which is based on figure 4 from Ref.~\cite{Aleynikov??}. 

When comparing figures~\ref{subfig:mysource} and~\ref{subfig:fullsource}, one is bound to notice the increased generation in the top left corner of figure~\ref{subfig:mysource}. 
This effect is due to the knock-on collisions of REs with finite pitch angles being able to generate secondary REs that have scattered to such an extent that they now move in the opposite direction of $-e\;E_\parallel$. 
These particles don't actually represent REs as the electric field will slow them down instead of accelerate them. 
Nevertheless, in the time it takes for these particles to slow down, they might still contribute to the avalanche via knock-on collisions. 
These particles are removed from the simulation when their energy drops below $\gamma_\text{min} m_e c^2$, at which point they can no longer contribute to the avalanche.

A further note to be made on the difference between a source term and a collision operator is that the collision operator can also be a sink in some regions of phase space. 
The knock-on collisions in general form a sink for the high energy REs with small pitch-angles and source for lower energy REs with higher pitch-angles. 
This matches reality, as the collisions do not only represent energy gain by the thermal bulk from the avalanche, but also an energy loss channel for the RE population to the bulk. 
This collision model, therefore, provides a way to accurately model the generation of trapped high energy electrons that, although they have no net gain of parallel velocity over a bounce period, we will refer to as trapped REs as they may become runaways due to detrapping, e.g., via knock-on collisions or via the effects of the Ware pinch~\cite{Ware1970, Nilsson_2015, Sarkimaki2022}.
Trapped REs are not subject to losses in stochastic fields to the same degree as freely moving particles are, such that a remaining population of trapped REs can act as an important seed for (re-)avalanching in case freely moving REs have largely been expelled, e.g., during the thermal quench of a disruption or during an RE termination event.
Capturing this part of the avalanche generation is made possible by removing markers from the simulation domain only when they've lost enough energy instead of removing them immediately. 

By choosing $\gamma_\text{min} < \gamma_c$, it is also possible to capture the effect of some of the REs dropping out of the runaway regime due to losing a significant amount of energy in a knock-on collision. 
From a computational standpoint, however, it should be noted that $\gamma_\text{min}$ should not be set too small to avoid overlap with the small angle collisions (that require a different numerical treatment) on the one hand and because of the rapid increase in computational cost as the knock-on collision time step gets on the order of the particle time step. 
To avoid this overlap with the small angle collisions, it should hold that $p_\text{min} =\sqrt{\gamma_\text{min}^2-1} \gg p_\text{th}.$

\subsection{Resampling}
\label{sec:resampling}

Energy and momentum conservation within the knock-on collisions come at the cost of tripling the number of markers each time the knock-on collisions are called upon. 
As most of the computational cost lies in the evolution of the markers, the tripling of markers makes this model computationally unfeasible to run for longer timescales, if nothing is done to decrease the number of markers again. 
If the marker population at any point exceeds $\frac{1}{3} N_\text{max}$, where $N_\text{max}$ is the upper limit for the number of markers for a given simulation, a resampling algorithm will be needed to decrease the number of markers before the knock-on collisions are called upon again. 
In this section we will discuss the resampling algorithm that has been implemented in JOREK for this purpose. 

The resampling is based on the GC model and will therefore be used to reproduce the phase space distribution $ f(\mathbf{X}, p_\parallel, \mu)$, where $\mathbf{X}$ is again the position of the guiding center, $p_\parallel$ the parallel momentum and $\mu$ the magnetic moment. 
This means that when the FO model is used, the markers are first converted to GC markers. 
For this conversion, the magnetic field $\mathbf{B}$ at the position of the guiding center is needed, which is not known before the conversion. Thus, an iterative approach is used, where a marker is converted first using the magnetic field at the FO position $\mathbf{x}$ and using the resulting GC position $\mathbf{X}$ to calculate $\mathbf{B}$ for the next iteration. 
This iteration converges quickly for most scenarios, but care needs to be taken in scenarios where $\mathbf{X}$ is ill defined. 
In such scenarios a lot of steps might be needed in the iteration to convert one particle, rapidly increasing the computational cost of the resampling, or, if the fields are very stochastic, the iteration might not converge at all anymore.
The resampling could then be extended relatively easily from the 3D+2P GC model domain to the 3D+3P FO model domain, but this comes at increased computational cost and is not necessary for the applications in this work, such that we use the conversion here and leave such algorithmic changes for future work.

After this conversion, the resampling consists of two separate steps. 
First, all markers will be binned spatially, i.e., each marker will be associated to some volume bin $V_j$ based on the position of their guiding center $\mathbf{X}_i$. 
Then for each volume bin $V_j$, the markers within $V_j$ will be sorted into momentum bins based on their $p_\parallel$ and $\mu$. 
We will first discuss the details of the spatial binning before moving on to the momentum binning. 

\begin{figure*}[ht!]
     \centering
     \begin{subfigure}[b]{0.32\textwidth}
         \centering
         \includegraphics[width=\linewidth]{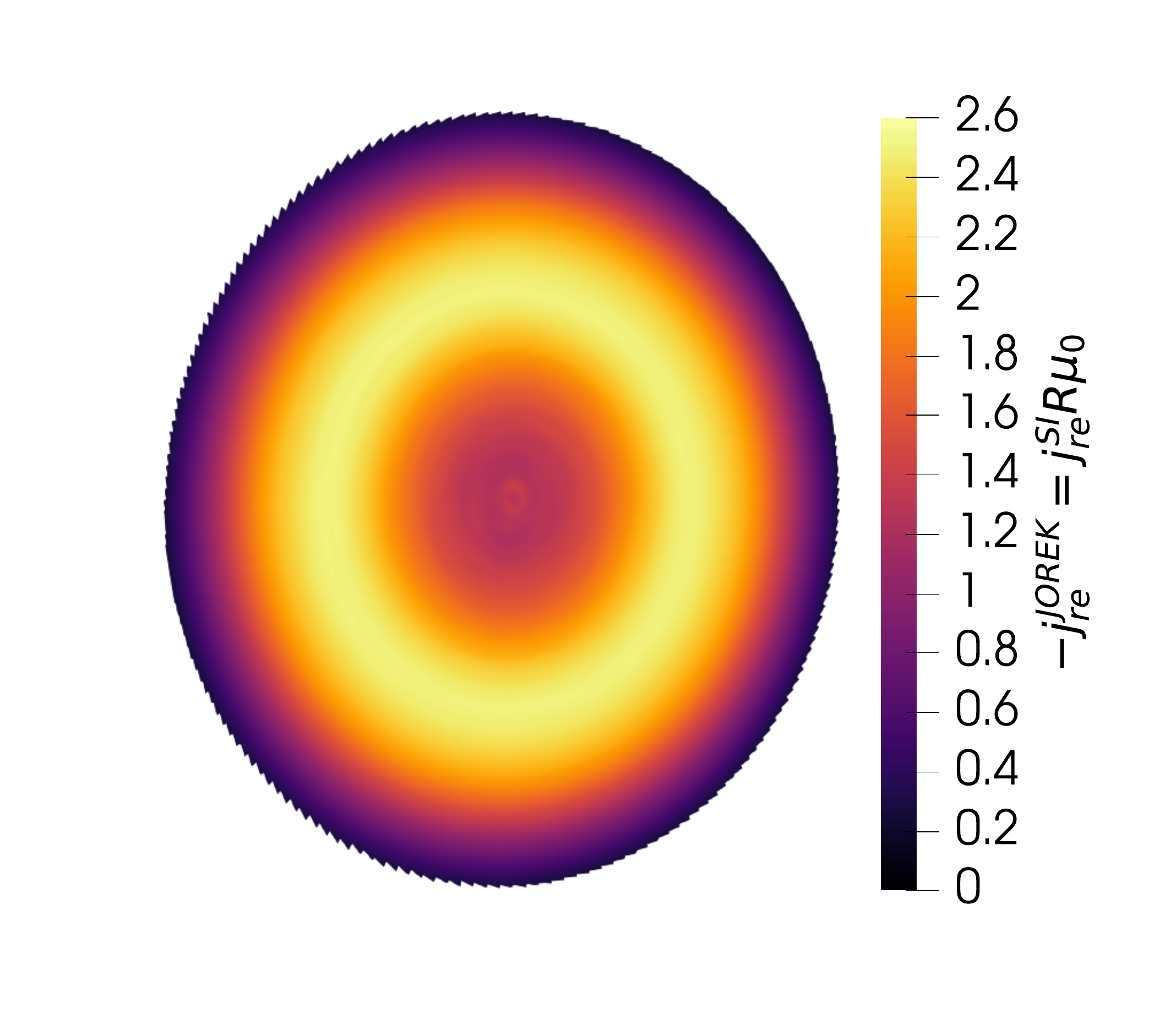}
         \caption{Without resampling}
         \label{subfig:beforeresampling}
     \end{subfigure}
     \hfill
     \begin{subfigure}[b]{0.32\textwidth}
         \centering
         \includegraphics[width=\linewidth]{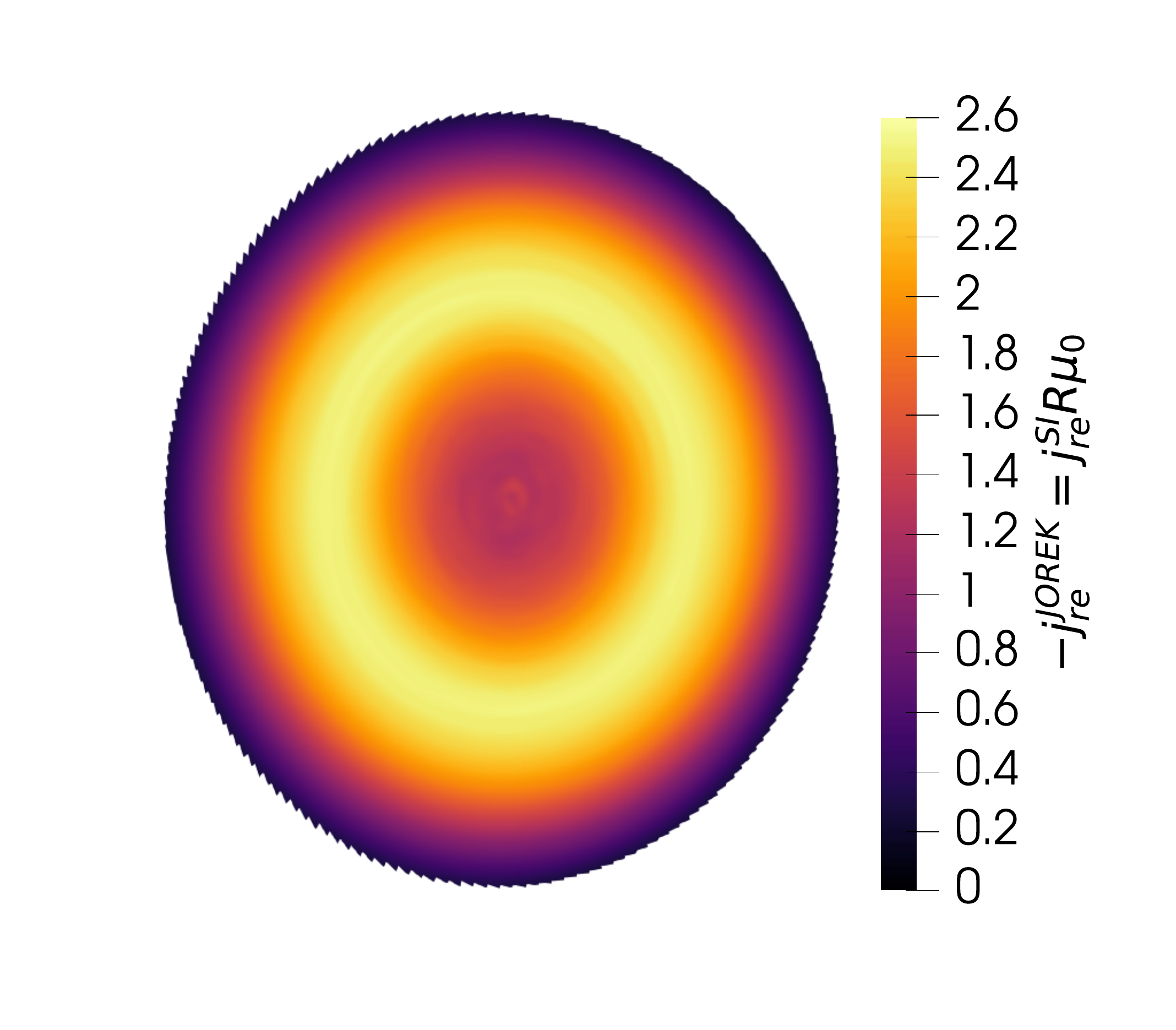}
         \caption{With resampling}
         \label{subfig:afterresampling}
     \end{subfigure}
     \hfill
     \begin{subfigure}[b]{0.32\textwidth}
         \centering
         \includegraphics[width=\linewidth]{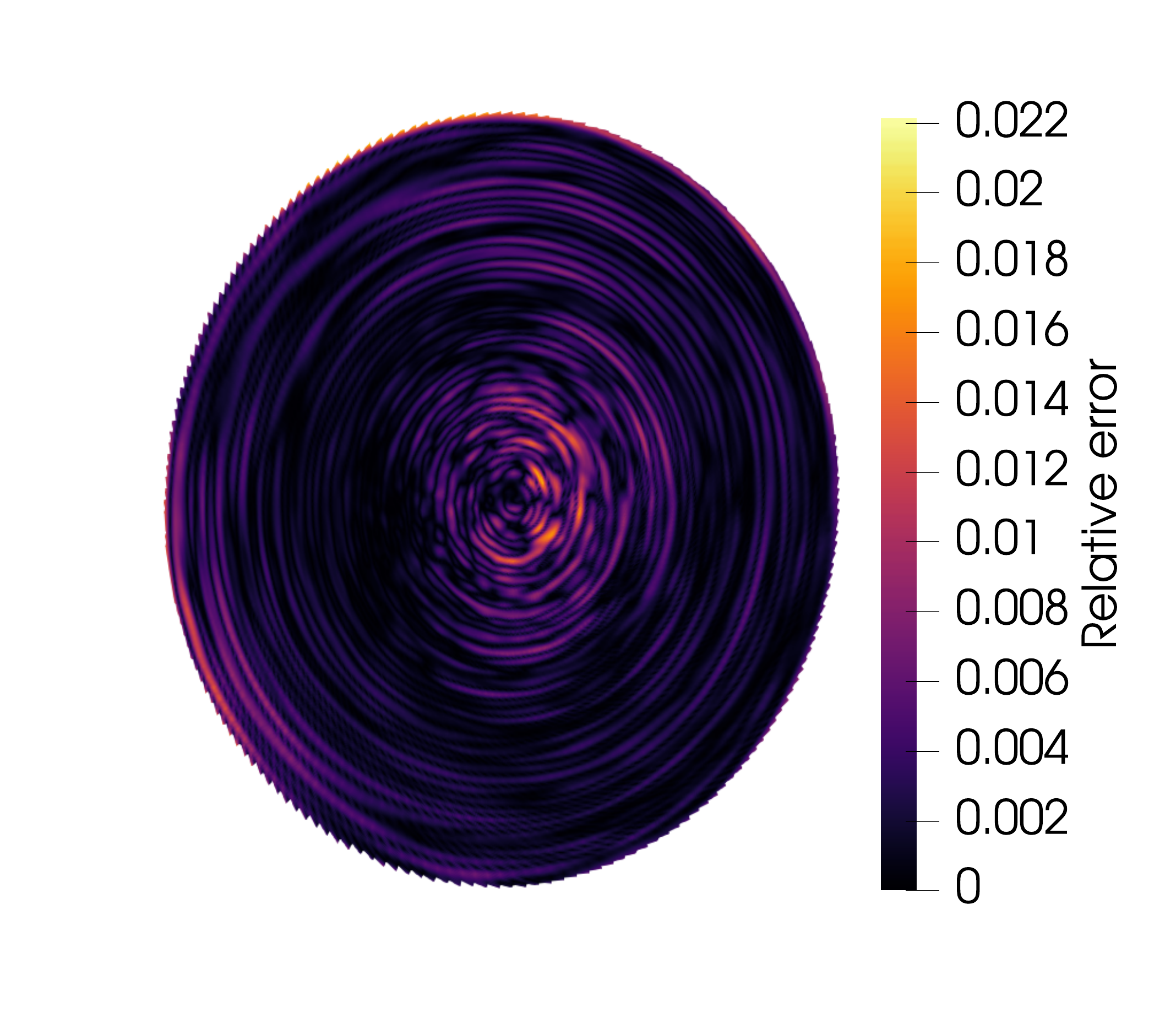}
         \caption{Relative error}
         \label{subfig:errorresampling}
     \end{subfigure}
        \caption{The effect of resampling on a hollow current profile is shown where the runaway current density is shown in (a) and (b) in JOREK units and the relative error is shown in (c). After the resampling, the number of markers is slightly less than a quarter of the original number of markers. In order to calculate the current density from the full-kinetic markers, parallel and hyper smoothing is used as well as averaging over one fluid timestep.}
        \label{fig:resampling_hollow}
\end{figure*}

\subsubsection{Spatial binning}
The grid that will be used for the spatial binning in the $R,Z$-plane is largely based on the finite element grid that is already used for the fluid part of the JOREK simulation. 
This choice is motivated by the fact that not only is it readily available, but this is also the grid the particle quantities will be projected onto for coupled simulations. 
Nevertheless, it might be that the resolution of the JOREK grid is lower than the ideal resolution for the resampling of the kinetic markers, in which case it is possible to further refine the resampling grid.  
To this end, the element-local coordinates ($s, t$) are used to divide the element into equal parts, which results in unequal parts in ($R, Z$) following the shaping of the grid (for more details on the JOREK grid and the element-local coordinates the interested reader is referred to Ref.~\cite{Hoelzl2021}). 
The 3D volume bin is then created using an equidistant binning method in the $\phi$-direction, using $n_\phi$ bins. 

The volume elements may have a large spread in the number of markers they contain. 
Some volume elements will be in regions with nested flux surfaces, where a lot of markers may be confined, while others will be inside stochastic regions or outside the separatrix, where few or no markers at all might be present. 
While resampling is expected to work well when there are many markers within a bin, the statistics will degrade when only a few markers are present. 
Therefore, we set a lower bound on the number of markers $m_\text{min}$ within $V_j$ such that when $m_j < m_\text{min}$ the markers within $V_j$ will not be resampled, but instead simply kept as they were. 
If, however, $m_j > m_\text{min}$ then all old markers $m_j$ will be removed and $\hat{m}_j < m_j$ new markers will be placed. 
These new markers are presently placed uniformly in $R, Z$ and $\phi$ within $V_j$ using a rejection sampling approach in $s, t$ coordinates based on the Jacobian of the transformation. 

The weight of the new markers in $V_j$ are all set equal to 
\begin{equation}
    \hat{w}^j = \frac{\sum_{i=1}^{m_j} w_i^j}{\hat{m}_j},
\end{equation}
such that the total weight $w_\text{tot}^j$ is conserved. 
The resampling thus conserves the RE density with the resolution of the spatial binning. 
The momenta $p_{\parallel,i}$ and $\mu_i$ of the new markers are set based on the momentum binning, as discussed in subsubsection~\ref{sec:mombin}.

To verify the applicability of the spatial binning, we will use it to investigate the effect of the resampling on the projection of the RE current density $j_\text{RE}$. 
For this, we will use states from the simulation of a beam termination scenario that has already been simulated with JOREK~\cite{Bergstroem2025}. 
As the markers are nearly mono-energetic in this particular simulation, the effect of the resampling will be mostly on the spatial distribution of the markers.

Hyper filters are used both in the direction of the magnetic field as well as in the $R-Z$ plane in order to project the RE current density $j_\text{RE}$ onto the finite element representation used in JOREK~\cite{Bergstroem2025, Vugt2019, Dvornova2021}.
These filters are shortly described in Appendix~\ref{A:filters}, but for more details the interested reader is referred to Refs.~\cite{Vugt2019, Dvornova2021}.
The numerical noise is then further decreased, helping to accurately resolve the physical processes~\cite{Lopez2025}, by taking the average of $j_\text{RE}$ over multiple particle timesteps (enough to make up one fluid timestep). 
Nevertheless, to create the smooth current profile that can be seen in figure~\ref{subfig:beforeresampling}, the number of markers used is $N_m^\text{initial} \approx 10^7$. 
To illustrate the effectiveness of the resampling for this case, we consider a case where the resampling reduces the number of markers to less than a quarter of its original value $N_m^\text{resampled} < \frac{1}{4} N_m^\text{initial}$. 
As for this simulation the resolution of the JOREK grid is sufficient, the volume bins for the resampling are created simply by using this grid and using $n_\phi = 18$ bins in the toroidal direction. 
Figure~\ref{fig:resampling_hollow} shows that in this case the resampled $j_\text{RE}$ closely resembles the non-resampled case.
The relative error between the two projections (see figure~\ref{subfig:errorresampling}) is naturally larger in places where $j_\text{RE}$ is lower, which can be expected as this is where less markers will be present. 
The error for the total current carried by the REs is in the order of $0.01 \%$. 

\begin{figure}[H]
     \centering
     \begin{subfigure}[b]{0.35\textwidth}
         \centering
         \includegraphics[width=\linewidth]{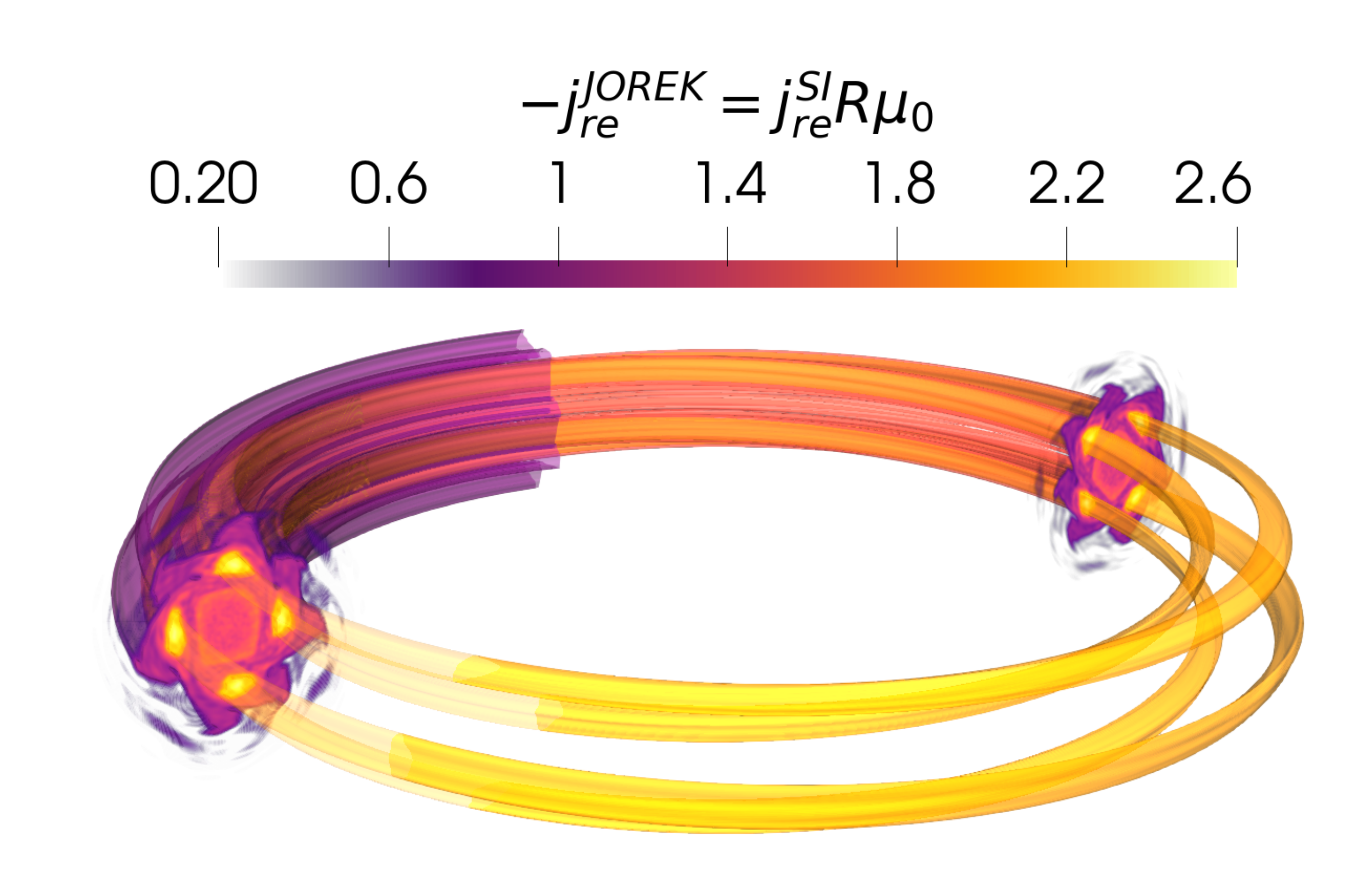}
         \caption{Without resampling}
         \label{subfig:3D_jre_nores}
     \end{subfigure}
     \hfill
     \begin{subfigure}[b]{0.35\textwidth}
         \centering
         \includegraphics[width=\linewidth]{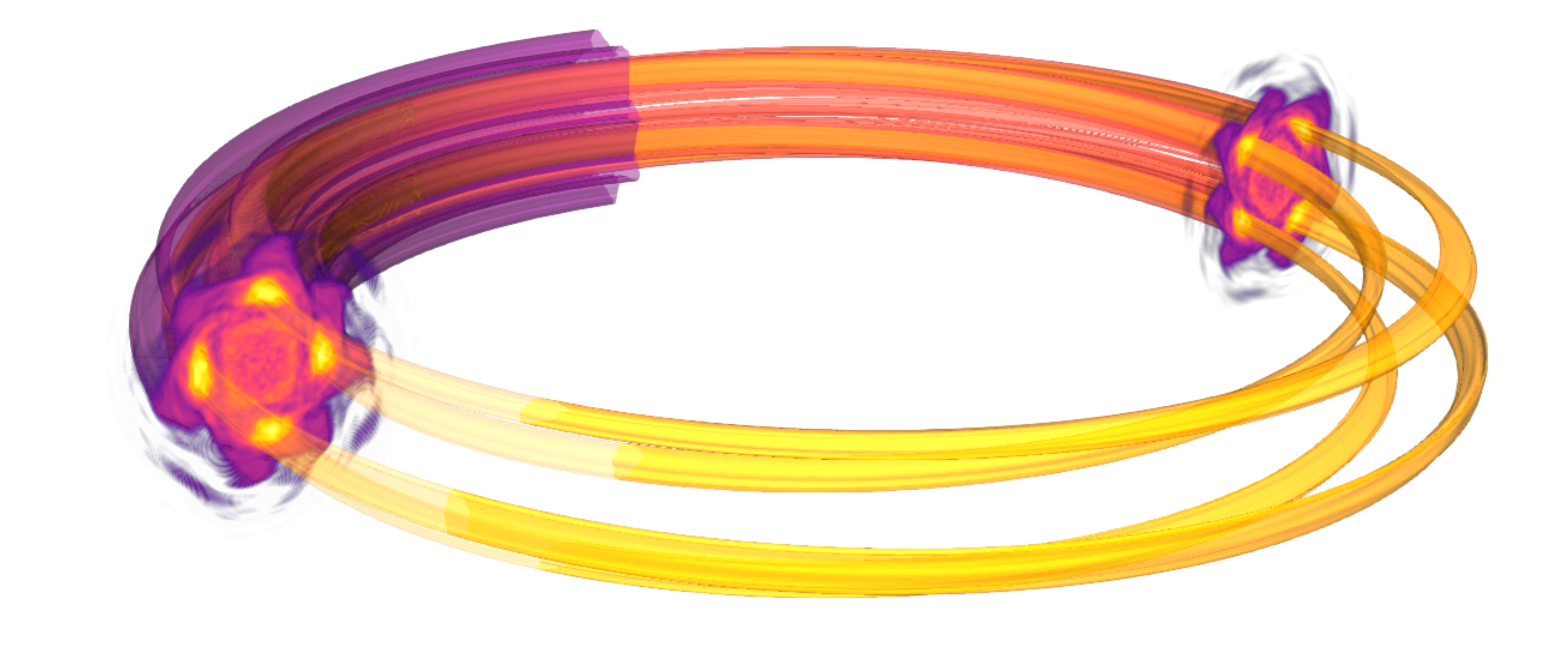}
         \caption{With resampling, $n_\phi = 18$}
         \label{subfig:3D_jre_18}
     \end{subfigure}
     \hfill
     \begin{subfigure}[b]{0.35\textwidth}
         \centering
         \includegraphics[width=\linewidth]{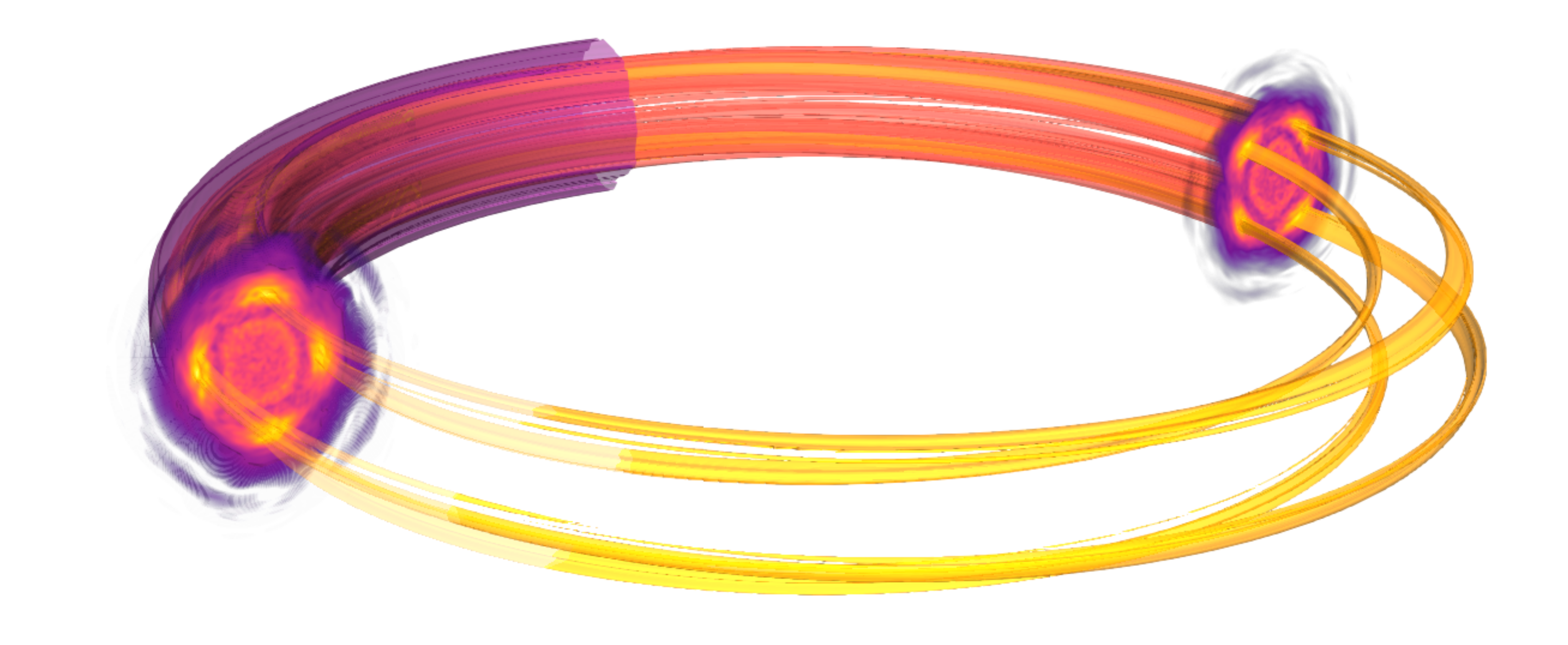}
         \caption{With resampling, $n_\phi = 3$}
         \label{subfig:3D_jre_3}
     \end{subfigure}
        \caption{The effect of the resampling on the 3D structure of the RE current density (in JOREK units) is illustrated by showing two poloidal cross-sections and using contours at equal values for all plots to illustrate how the islands wrap around the torus. Comparing the resampled distribution (b) with the initial distribution (a) shows that the 3D structure is well preserved. When $n_\phi$ is chosen too small (as in (c)), however, part of the 3D structure is lost even though the island structure still remains visible.}
        \label{fig:3D_jre}
\end{figure}

Next, we will look at a scenario where the toroidal binning is necessary to resolve a 3D RE distribution in the presence of a magnetic island structure. 
To visualize this, we consider the $j_\text{RE}$ at a later point in the simulation where the island structure is clearly visible, as illustrated in figure~\ref{fig:3D_jre}.
Here not only a poloidal cross-section of the current profile can be seen, but also contour plots at given values to show the extent of some of the structures in the toroidal direction. 
Figure~\ref{subfig:3D_jre_3} shows that if too few toroidal bins are used, the resampling is unable to resolve accurately how the island structure twists helically around in the toroidal direction and the current density gets smeared poloidally. 
However, when one uses a sufficient number of toroidal bins, as done in figure~\ref{subfig:3D_jre_18}, the island structure remains clearly visible.
This means that an informed decision must be taken when $n_\phi$ is set at the beginning of the simulation, or should be adjusted throughout the simulation based on which modes are present.

\begin{figure*}[p]
    \centering
    \includegraphics[width=0.9\linewidth]{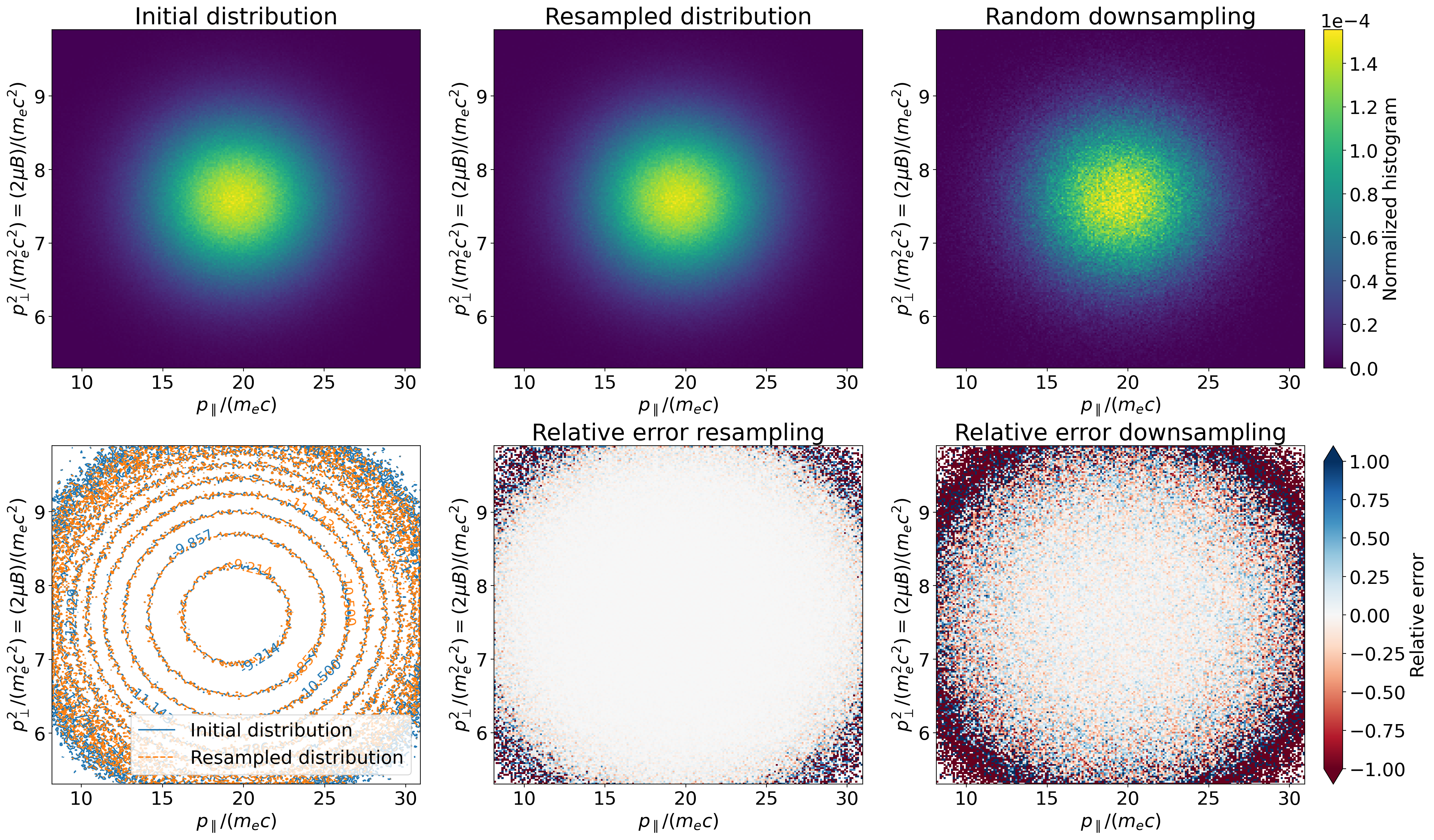}
    \caption{The effect of resampling in momentum space is tested on a Gaussian distribution where $10^7$ markers are used for the initial distribution and only $3\times10^6$ for the resampled and random downsampled distributions. Top row shows histograms of the PiC data before and after the resampling created using 200 bins in both the $p_\parallel$ and $\mu$ direction and are equal to the bins used for the resampling. A contour plot of the resampled distribution on top of the original distribution is shown on the bottom left. The relative errors caused by the resampling and by a naive random downsampling method are shown on the bottom center and right. The resampling is able to reproduce the initial distribution with less error than the random downsampling. }
    \label{fig:Nres_sob}
         \centering
         \hfill
     \begin{subfigure}[b]{0.45\textwidth}
         \centering
         \includegraphics[width=\linewidth]{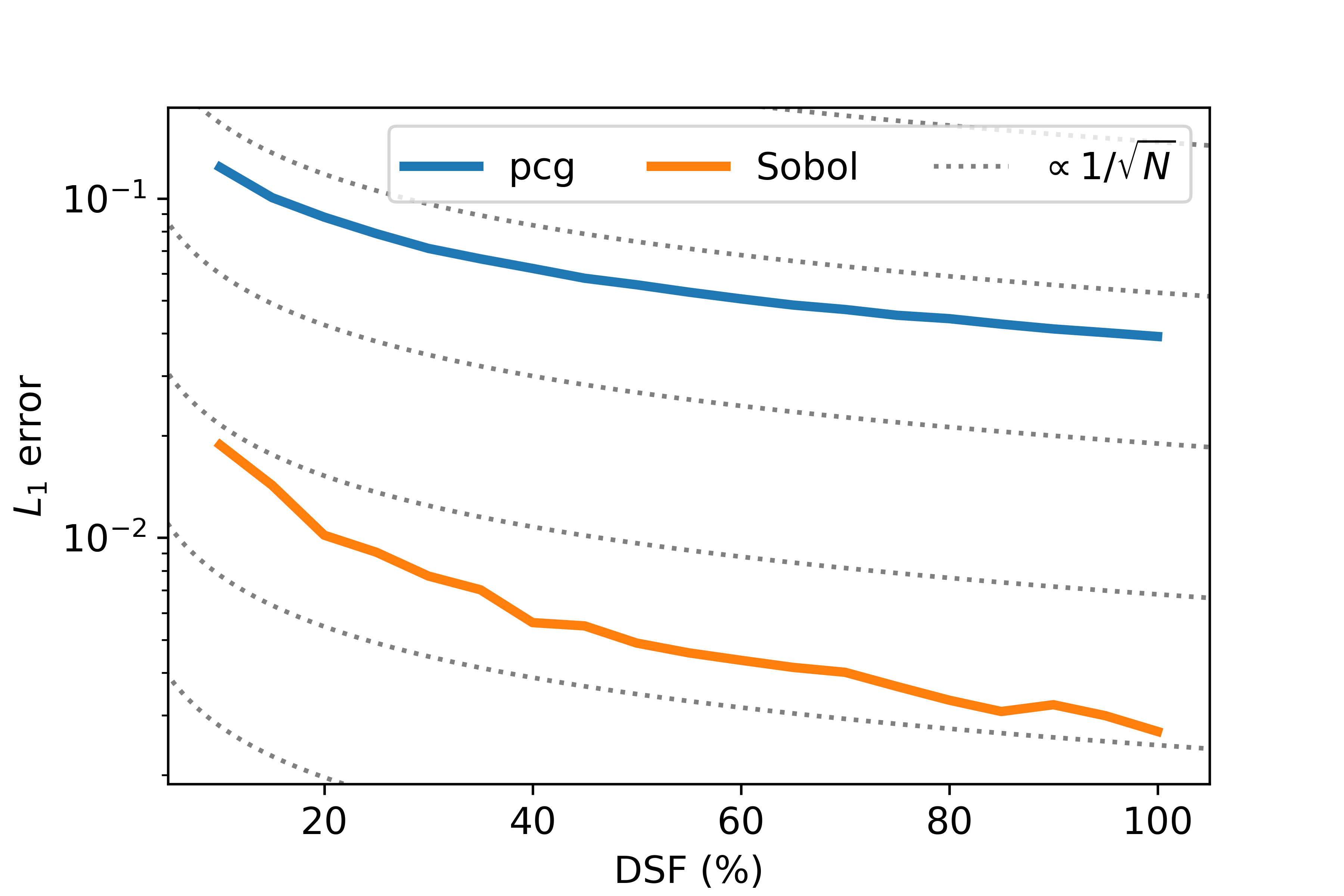}
         \caption{$L_1$ error}
         \label{subfig:L1}
     \end{subfigure}
     \hfill
     \begin{subfigure}[b]{0.45\textwidth}
         \centering
         \includegraphics[width=\linewidth]{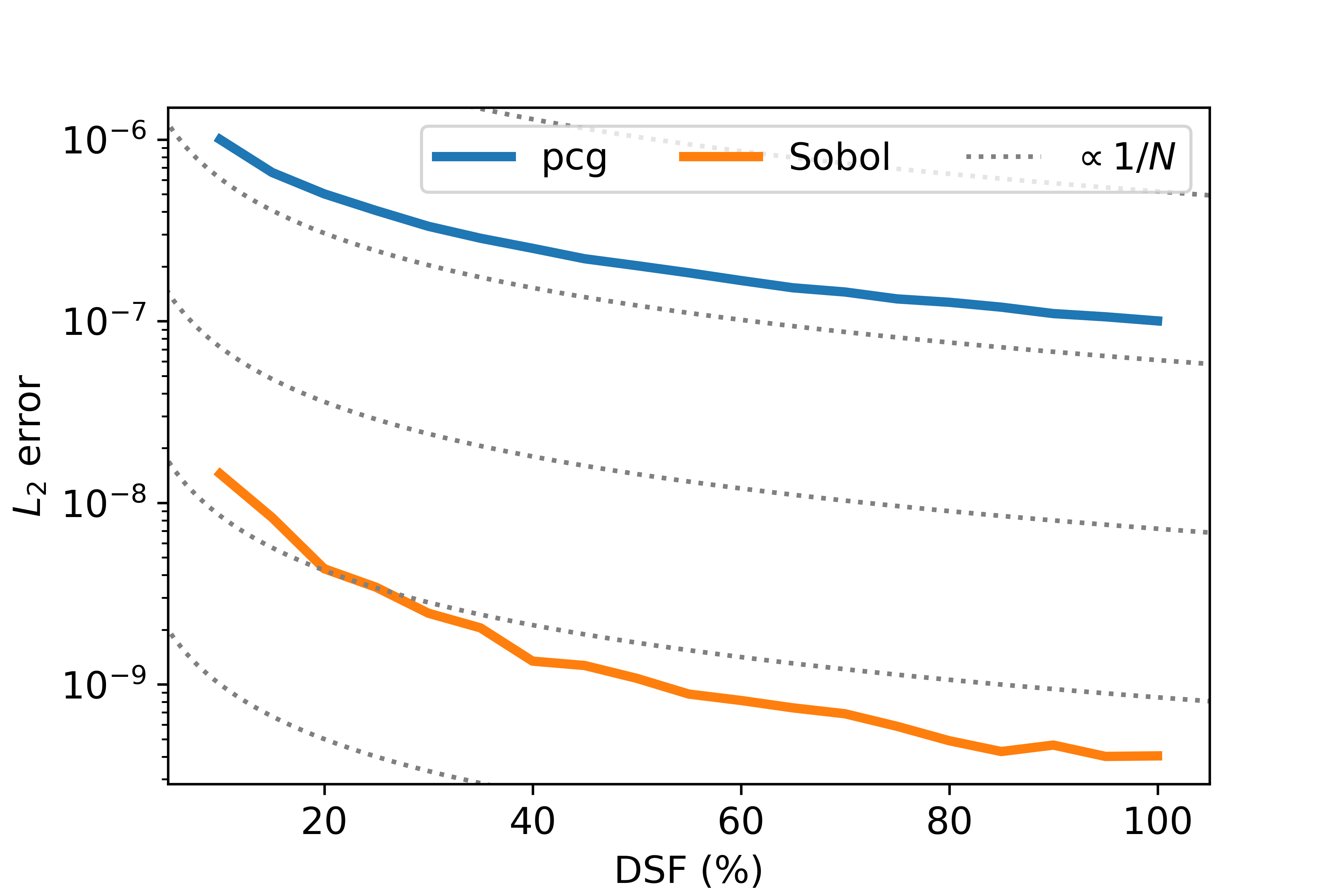}
         \caption{$L_2$ error}
         \label{subfig:L2}
     \end{subfigure}
     \hfill
        \nextfloat \caption{The $L_1$ (left) and $L_2$ (right) error scalings as a function of the down sampling factor (DSF) are shown for the momentum space resampling both using the methods pcg rng and the Sobol sequence. The resampling used 200 bins in both the $p_\parallel$ and $\mu$ directions. The dotted lines illustrate the scaling. The Sobol sequence both allows for lower errors and scales better with the DSF.}
        \label{fig:error_N}
\end{figure*}

\subsubsection{Momentum binning}
\label{sec:mombin}
Within each volume element, the markers are binned equidistantly based on their $p_\parallel$ and $\mu$. 
Although the same number of momentum space bins is used for each $V_j$, the upper and lower bounds are determined for each volume bin separately based on the maximum and minimum values of $p_\parallel$ and $\mu$ of the markers contained in it. 
More precisely, the bounds for volume bin $j$ are set as
\begin{align}
    p_{\parallel,j}^\text{upper bound} &= p_{\parallel, j}^\text{max} + \frac{p_{\parallel, j}^\text{max} - p_{\parallel, j}^\text{min}}{2 (\sqrt{N_\text{bins}} - 1)}, \\
    p_{\parallel,j}^\text{lower bound} &= p_{\parallel, j}^\text{min} - \frac{p_{\parallel, j}^\text{max} - p_{\parallel, j}^\text{min}}{2 (\sqrt{N_\text{bins}} - 1)}, \\
    \mu^\text{upper bound}_j &= \mu_{j}^\text{max} + \frac{\mu_{j}^\text{max} - \mu_{j}^\text{min}}{2 (\sqrt{N_\text{bins}} - 1)} \text{ and}\\
    \mu^\text{lower bound}_j &= \mu_{j}^\text{min} - \frac{\mu_{j}^\text{max} - \mu_{j}^\text{min}}{2 (\sqrt{N_\text{bins}} - 1)},
\end{align}
where $N_\text{bins}$ is the number of bins used for the momentum binning, such that the center of the outermost bins corresponds to $p_\parallel^\text{max}$, $p_\parallel^\text{min}$, $\mu^\text{max}$ and $\mu^\text{min}$.

After the markers have been binned, the result is used to calculate a probability $P_i^j$ for a new marker within $V_j$ to have $p_{\parallel}$ and $\mu$ corresponding to values within momentum bin $i$. 
For each marker, a momentum bin is selected based on this probability using inversion sampling. 
Once a momentum bin has been selected, the new marker will be given a $p_{\parallel}$ and $\mu$ corresponding to a value chosen uniformly within momentum bin $i$. 
This method will not allow for the conservation of energy or momentum within the volume bin, but will allow for the reproduction of the momentum space distribution.
In figure~\ref{fig:Nres_sob} the resampling was tested on a Gaussian distribution and a comparison with a naive random downsampling method, where a random sample is selected and removed after which the weights of the remaining markers are rescaled to conserve the total weight, is shown. 

Before moving on to the error scaling of the resampling, we will first make a slight detour into the types of random number generators that are readily available in JOREK. 
For the particles module in JOREK, a lot of random numbers are used for different purposes and it is important that they are both of good quality and can be generated on multiple threads. 
This means that the built-in generator for different Fortran compilers is not always suitable, which is why pseudorandom and quasirandom number generators (rngs) are used in JOREK (see Ref.~\cite{Vugt2019} for details). 
The pseudorandom number generator is of the permuted congruential generator (pcg)~\cite{Oneill2014} family and is the easiest one to use as it does not have restrictions on the number of threads, unlike the quasirandom number generator based on the Sobol sequence~\cite{Sobol1967} where the number of threads must be a power of 2. 
If MPI parallelization is used as well, as is done for the resampling, the number of tasks also needs to be a power of 2. 
Even though the Sobol sequence gives the possibility to achieve better convergence than the $1/\sqrt{N}$ obtained using the pcg rng~\cite{NumericalRecipes}, the resampling was implemented such that the user can easily choose between them to allow for usage of the resampling without added restrictions. 

The effect of the difference between the pcg rng and the Sobol sequence is made clear when the error scaling of the resampling with the downsampling factor (DSF) is considered, where the DSF is the number of new markers as a percentage of the number of old markers. 
Figure~\ref{fig:error_N} shows that both the $L_1$ (sum of the absolute difference between bins) and $L_2$ (sum of the squared difference between bins) errors scale well with the DSF. 
The dotted lines show the expected scaling for a pseudorandom rng and it is clear that in both cases this scaling is matched when the pcg rng is used. 
The Sobol sequence scales better in both cases than the pcg rng, even though the most optimal scaling of $1/N$ is not achieved (it approaches the soft limit of $N^{-\frac{2}{3}}$ for high numbers of markers). 
The relative errors in $\bar{p}_\parallel$ and $\bar{\mu}$ are in the order of $10^{-5}$ and $10^{-7}-10^{-6}$ when using the Sobol sequence and $10^{-4}$ for both when using the pcg rng. 
For this case the relative error in the total energy is in the order of $10^{-6}$ when using the Sobol sequence and $10^{-4}$ when using the pcg rng.

\begin{figure}[H]
    \centering
    \includegraphics[width=.8\linewidth]{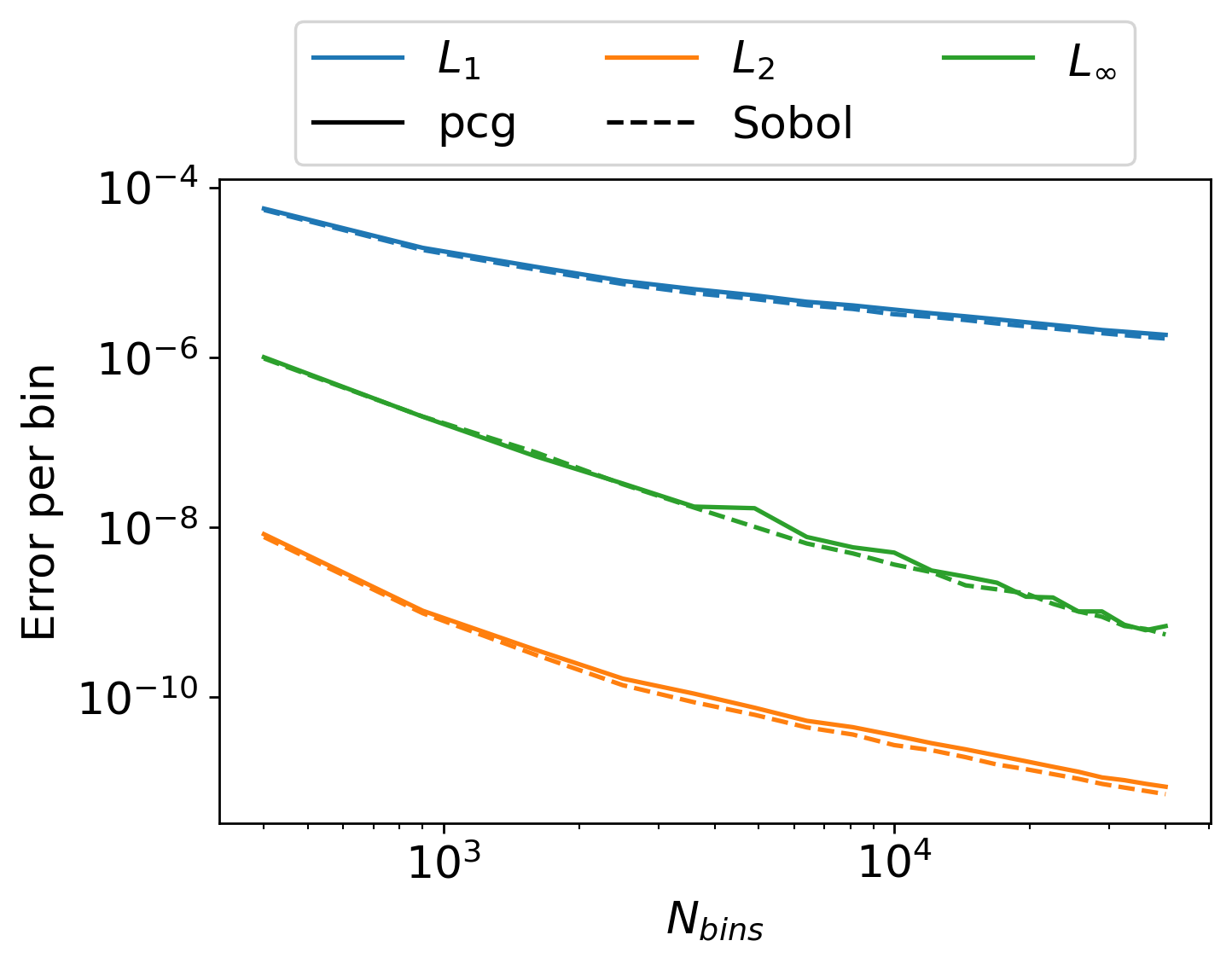}
    \caption{The error per bin in momentum space, for varying number of bins used for the resampling. Resampling was done on an initial Gaussian test distribution of $10^7$ markers, which were resampled to $\frac{1}{3} \times 10^7$ markers.}
    \label{fig:error_per_bin}
\end{figure}

The error introduced by the resampling depends not only on the number of markers and random number generator used, but also on the number of bins used for the momentum binning. 
Figures~\ref{fig:Nres_sob} and~\ref{fig:error_N} have been created using 200 bins in both $p_\parallel$ and $\mu$, i.e., with $N_\text{bins} = 40000$ bins in total. 
It is shown in figure~\ref{fig:error_per_bin} that the $L_1$, $L_2$ and $L_\infty$ (maximum absolute difference between bins) errors indeed decrease with increasing $N_\text{bins}$ as expected and that the decrease is roughly independent of the random number generator used.

\subsubsection{Usage of the resampling}
Up to this point we described the algorithms and implementation of the resampling and demonstrated applications to specific test cases. We will be using the resampling exactly in the described manner for section~\ref{sec:application}.

For demanding production applications, further optimizations might be needed. We discuss some of these possible adaptations in the following, both covering the computational side and the applicability to different scenarios and adaptability.
In particular, also an example of applying the resampling to the avalanche distribution will be shown.

The resampling is capable of generating a new set of markers that approximates the same 5D distribution function as the original markers. In practice, the accuracy requirements for the respective application set an lower limit for the number of markers after the resampling.
As the knock-on collisions triple the number of markers, the resampling typically needs to be applied every 1-3 times the knock-on collisions is called upon. This limit depends on the maximum number of markers acceptable for computational reasons and the fraction of markers that may be lost in between knock-on collisions. 
Even though the computational cost of the resampling exceeds that of the knock-on collisions, it is still significantly lower than the particle pushing. 
The resampling is, therefore, not a limiting factor on the computational cost of the simulation.

The resampling described above relies on linear bins in both $p_\parallel$ and $\mu$. If specific regions of phase space need to be resolved with particularly high accuracy, this may need to be adjusted, which the implementation easily allows for. 

\begin{figure*}[p]
    \centering
    \includegraphics[width=0.9\linewidth]{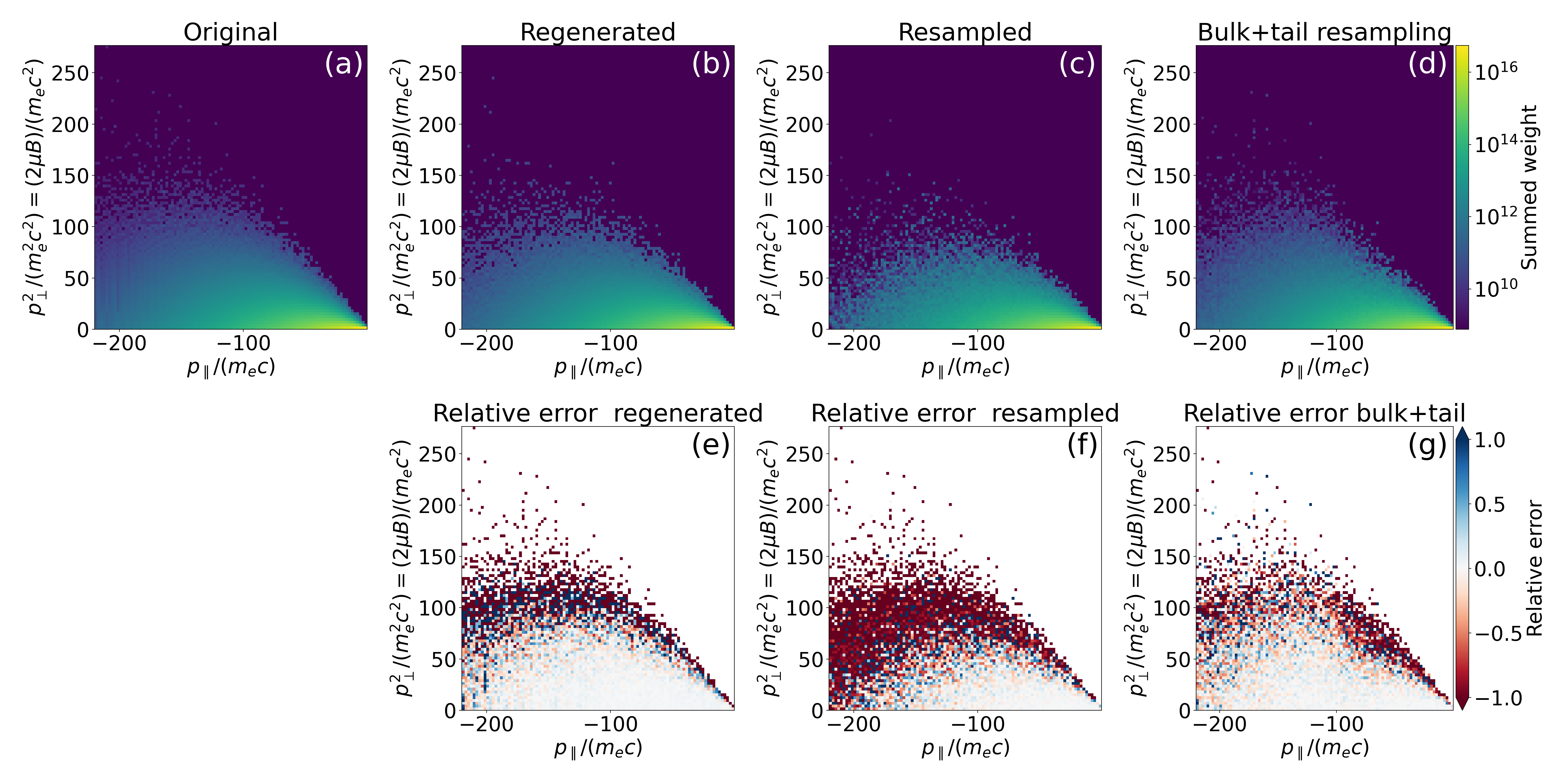}
    \caption{Distribution functions for the resampling tests conducted based on an avalanche distribution function. The figures in the top row show a histogram in momentum space of all markers: a) corresponds to the original distribution with $10^7$ markers created with priority sampling, b) to priority sampling with $2.3\times10^6$ markers, c) to the distribution after resampling from $10^7$ to $2.3\times10^6$ markers with equally treated momentum bins, and d) to the distribution after resampling with adjusted momentum binning.
    The bottom row of figures shows the relative error between each distribution from the upper row and the initial distribution.}
    \label{fig:avares_momentum}
    \centering
    \hfill
     \begin{subfigure}[b]{0.20\textwidth}
         \centering
         \includegraphics[height=0.17\textheight]{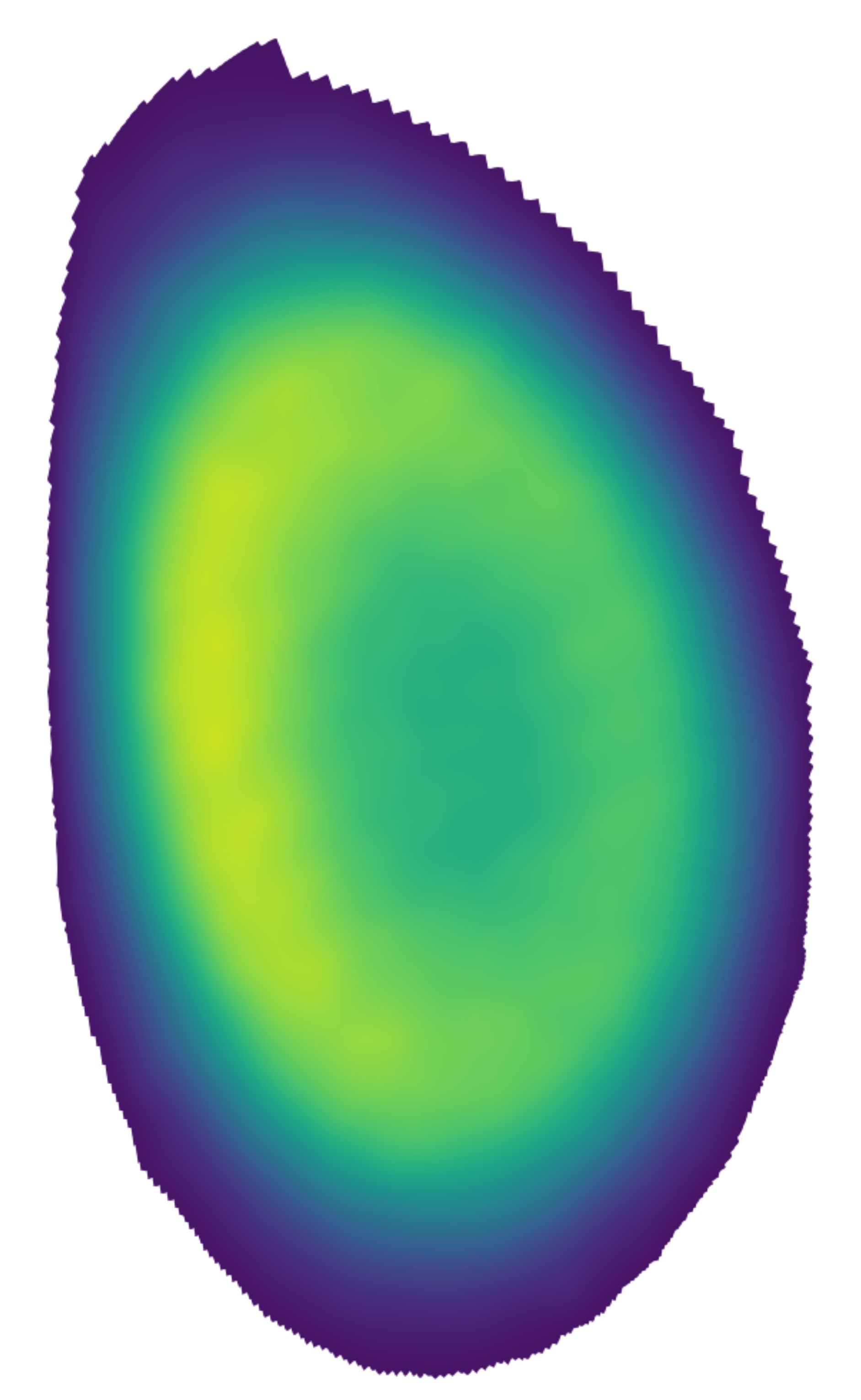}
         \caption{Initial}
     \end{subfigure}
     \begin{subfigure}[b]{0.20\textwidth}
         \centering
         \includegraphics[height=0.17\textheight]{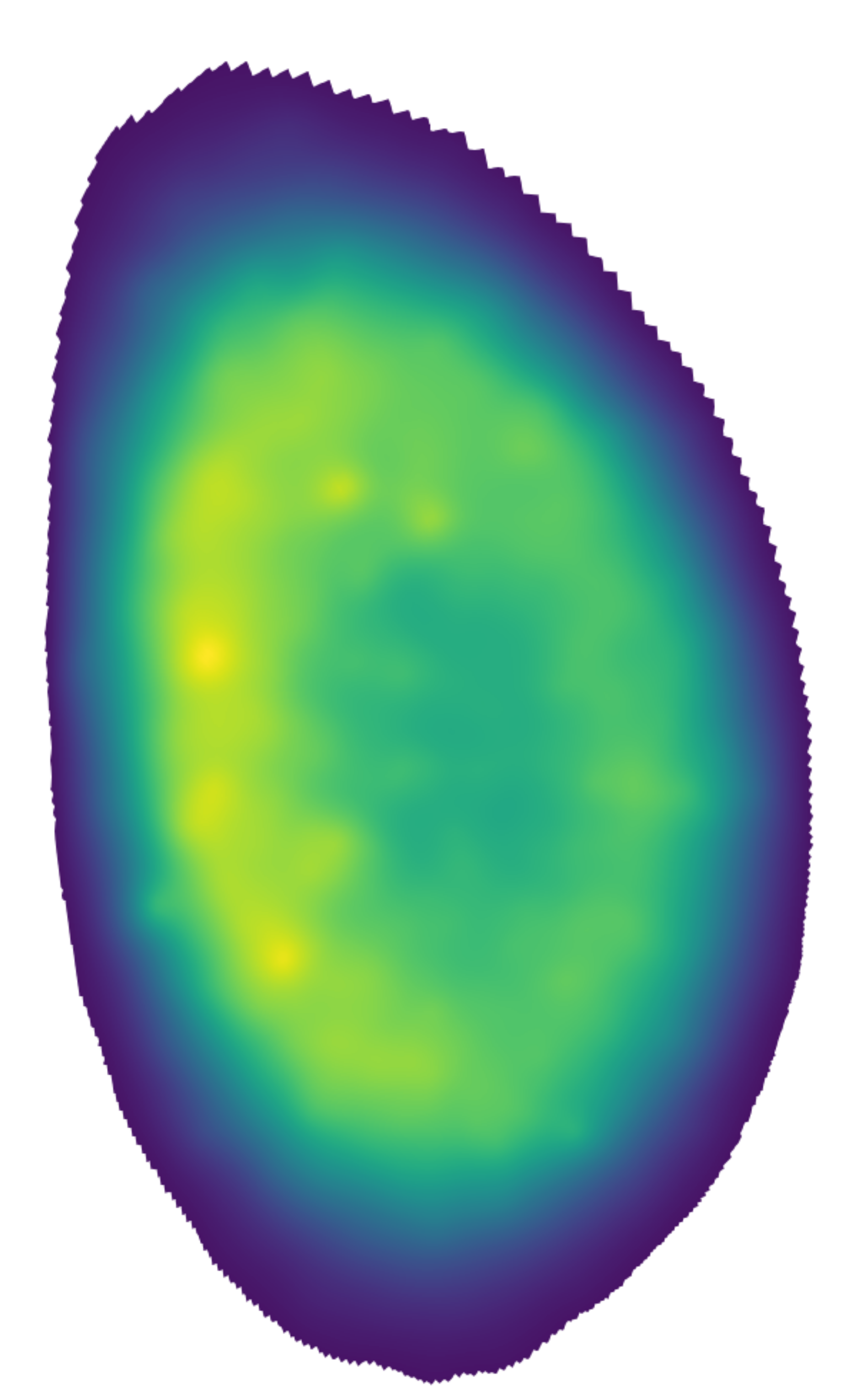}
         \caption{Regenerated}
     \end{subfigure}
     \begin{subfigure}[b]{0.20\textwidth}
         \centering
         \includegraphics[height=0.17\textheight]{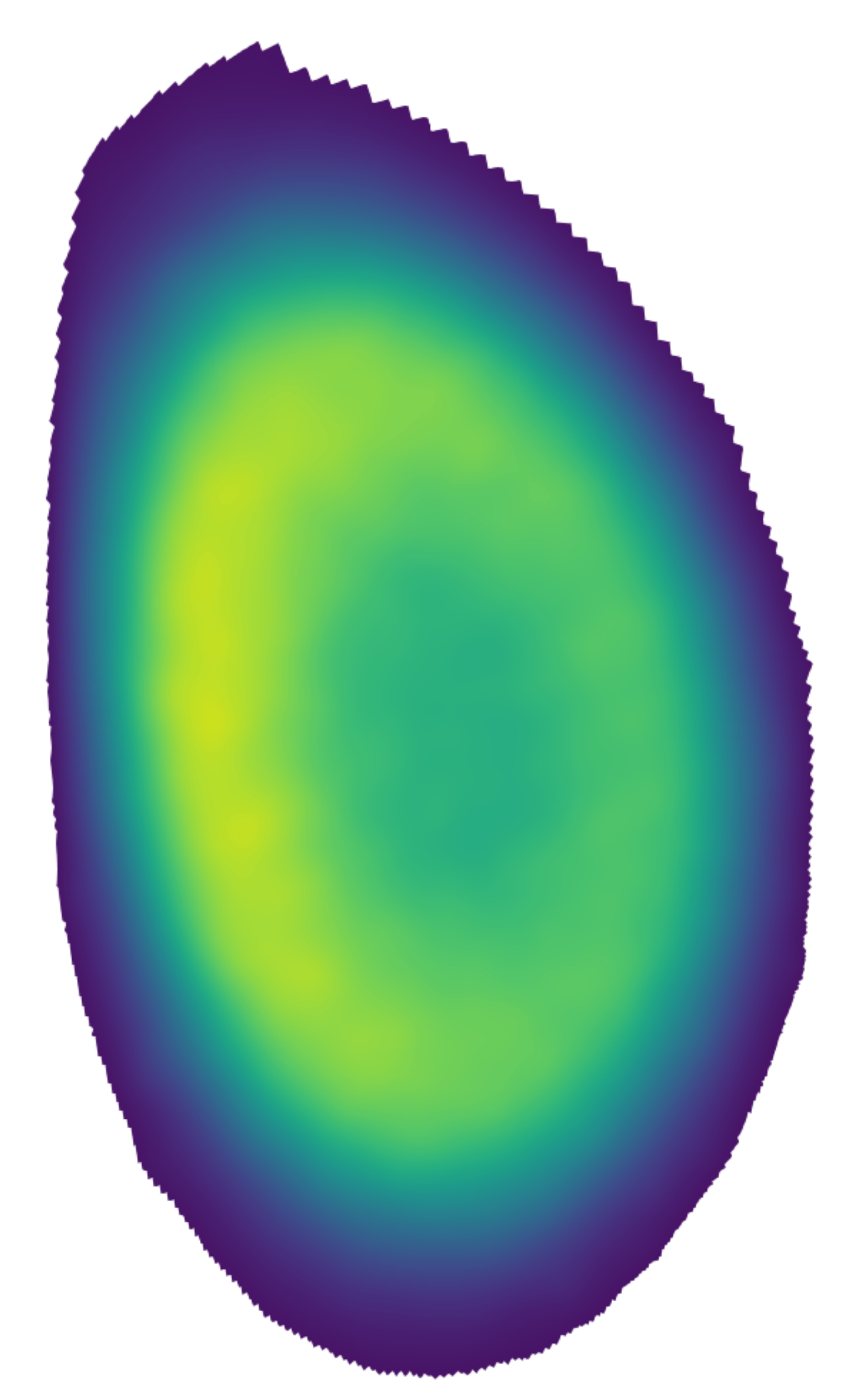}
         \caption{Resampled}
     \end{subfigure}
     \begin{subfigure}[b]{0.28\textwidth}
         \centering
         \includegraphics[height=0.17\textheight]{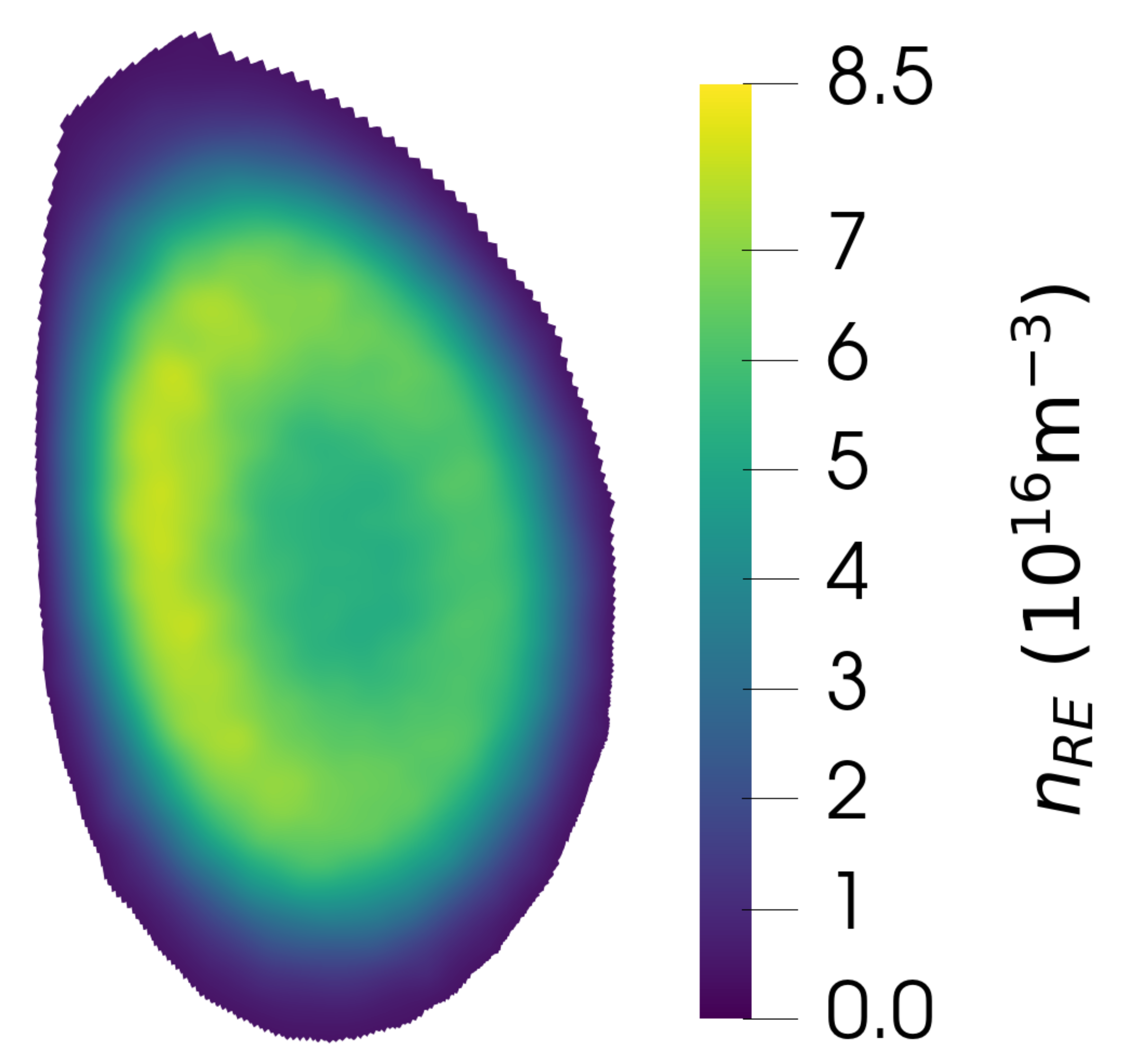}
         \caption{Bulk+tail resampled}
     \end{subfigure}
    \hfill
          
     \hfill \begin{subfigure}[b]{0.20\textwidth}
         \centering
         \includegraphics[height=0.17\textheight]{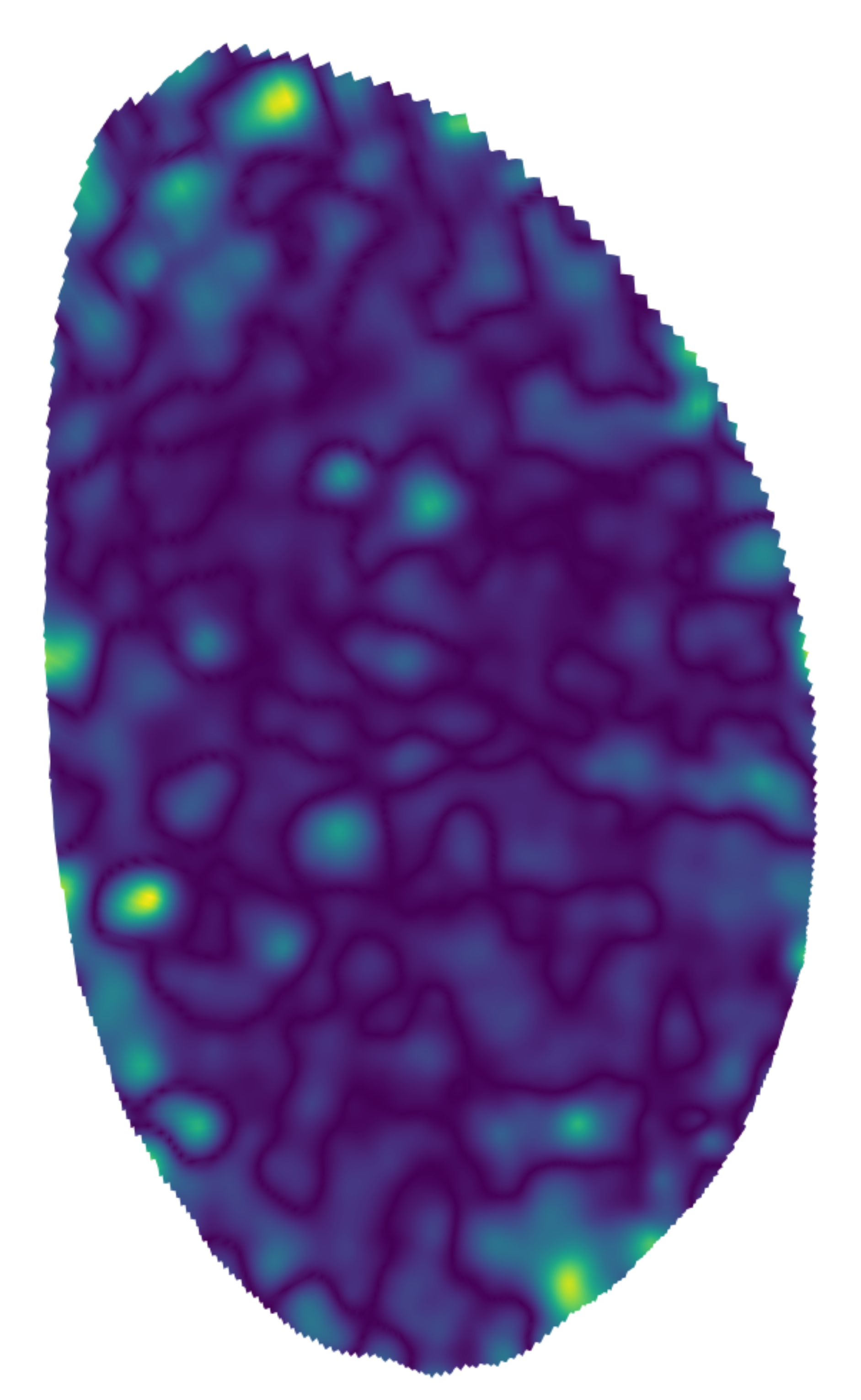}
         \caption{Regenerated}
     \end{subfigure}
     \begin{subfigure}[b]{0.20\textwidth}
         \centering
         \includegraphics[height=0.17\textheight]{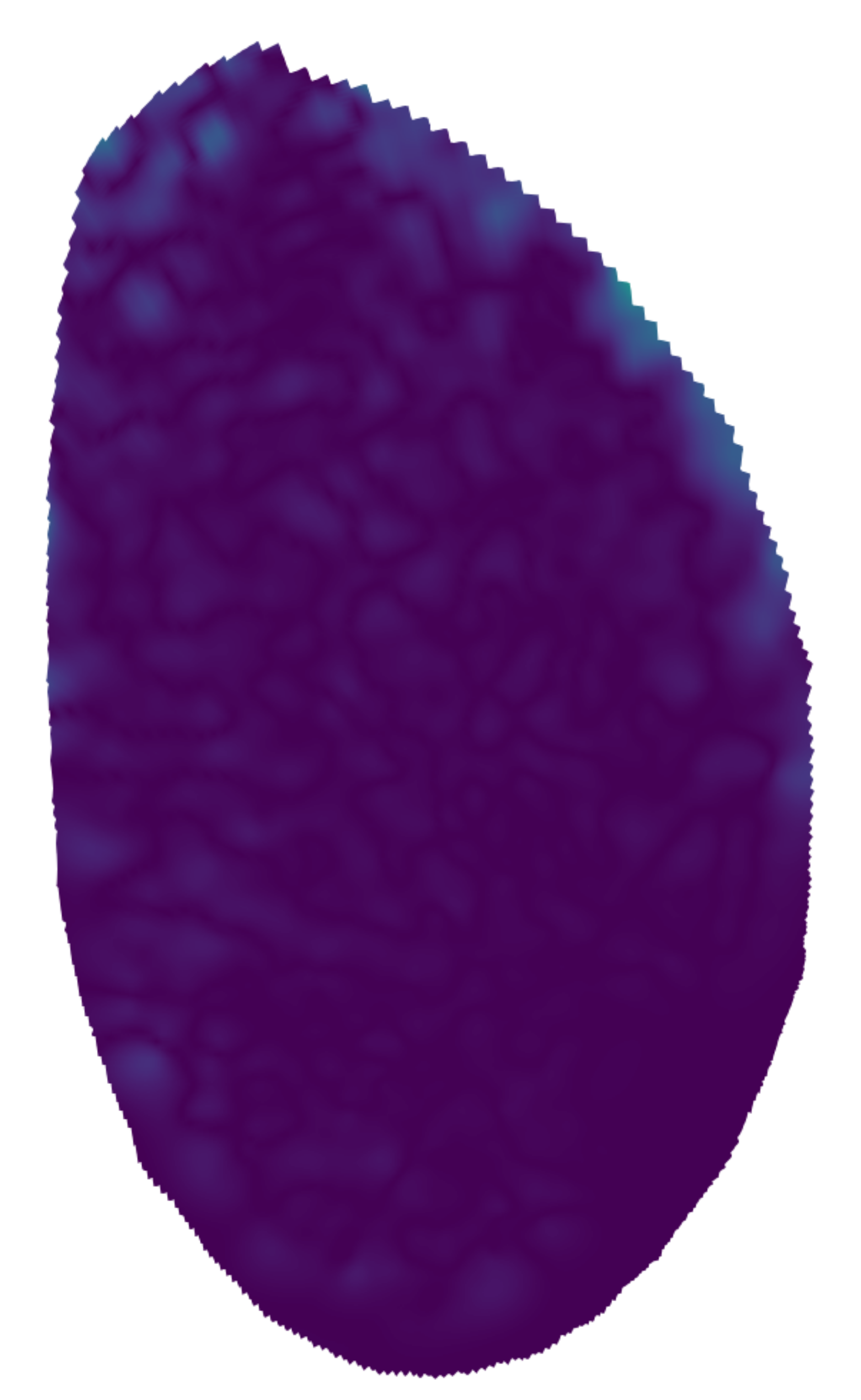}
         \caption{Resampled}
     \end{subfigure}
     \begin{subfigure}[b]{0.28\textwidth}
         \centering
         \includegraphics[height=0.17\textheight]{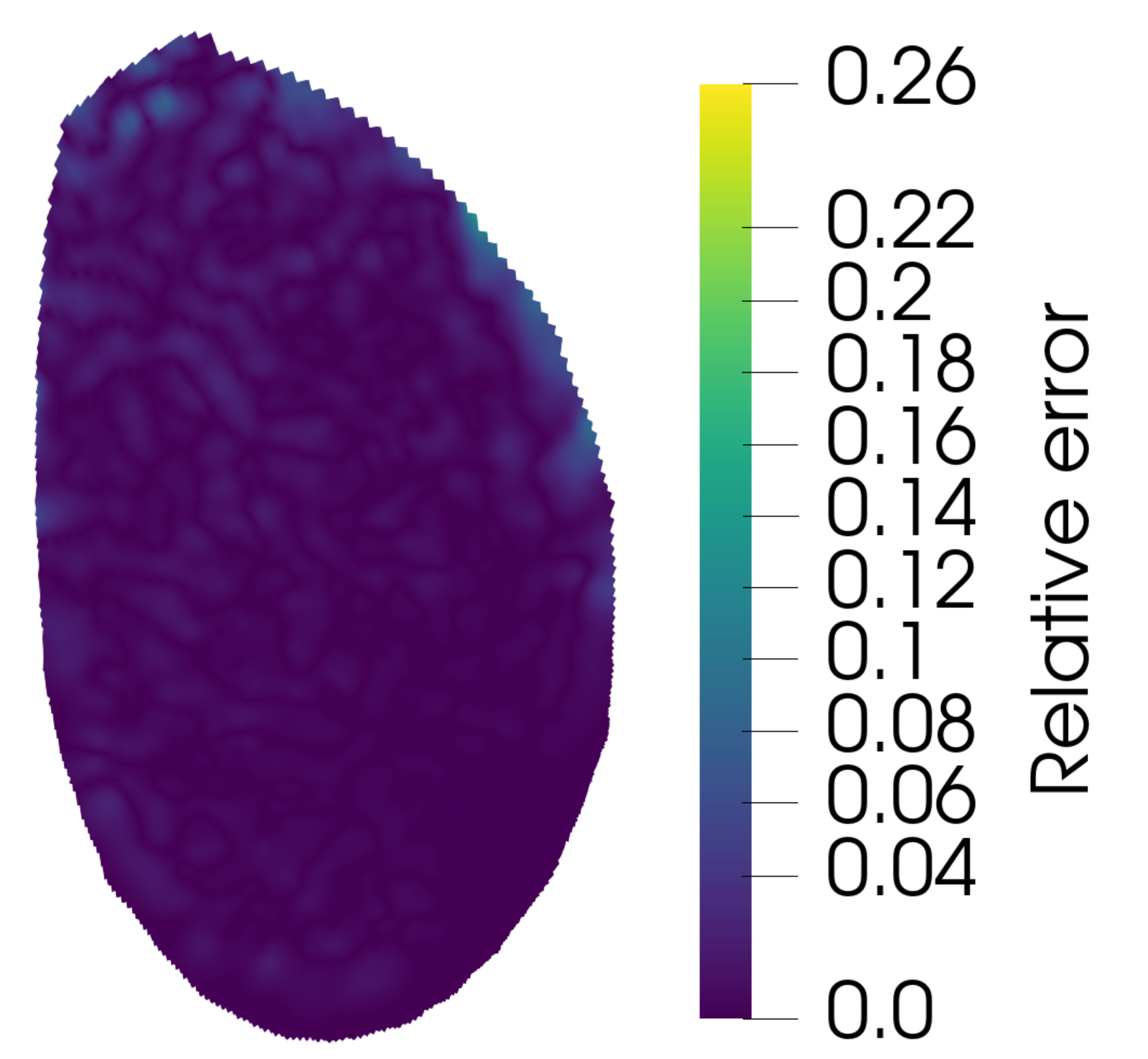}
         \caption{Bulk+tail resampled}
     \end{subfigure}
        \nextfloat\caption{The runaway electron number density is shown in figures (a-d) for the initial distribution (where $N=10^7$) and the regenerated, resampled and bulk+tail resampled distributions (where $N\approx 2.3\times10^6$). The relative error of these last three with respect to the initial distribution is shown in (e-g).}
        \label{fig:nre_avares}
\end{figure*}

To demonstrate this, we will consider one challenging application to a strongly varying phase space distribution. 
For this purpose we will use the analytical quasi-steady-state runaway distribution of REs in the scenario that the avalanche source dominates the runaway generation. 
In a spatially uniform and fully ionized plasma with a constant electric field, this distribution is given by \cite{Embreus2018b, Fulop2006}
\begin{align}
    f (p, \xi, t) &= \frac{n_\text{RE}(t) A(p)}{2 \pi m_e c p^2 c_Z \ln(\Lambda)} \frac{\exp\Bigl[-\frac{\gamma}{c_Z \ln(\Lambda)} - A(p)(1+\xi) \Bigr]}{1-e^{-2A}}, \\
    A(p) &= \frac{E/E_c +1}{Z_\text{tot}+1} \gamma, \\
    n_\text{RE}(t) &= n_\text{RE} (0) e^{t/t_\text{ava}}, \\
    t_\text{ava} &= \frac{c_Z}{4 \pi n_\text{tot} r_e^2 c (E/E_c -1)}, 
\end{align}
where $c_Z \approx \sqrt{5 + Z_\text{eff}}$ with the effective charge $Z_\text{eff} = \sum_i n_i Z_i^2/\sum_i n_i Z_i$ with the sum taken over all ion species $i$. 
In order to create a test distribution for the resampling, we sample markers from this distribution, while making sure enough markers are used to represent the tail and adjust the spatial distribution according to a preset runway electron density from a fluid simulation. 
The weights of the markers are rescaled in order to match the total RE density, while ensuring that they still represent the analytical distribution. 
Since we are creating this 5D distribution purely to show the effect of the resampling on it, the case was chosen arbitrarily and does not describe a realistic experimental scenario. 

The initial distribution was sampled using $1\times 10^7$ markers and a priority distribution that increases the number of markers in the tail and reduces their weights accordingly to ensure that this phase space domain is resolved accurately. When considering resampling to $2.3 \times 10^6$ markers with equally treated momentum bins and no priority sampling, the method fails to capture the details of the high energy tail of the distribution as seen in Figure \ref{fig:avares_momentum}.

To improve the results, an adjustment is made to the momentum space binning as follows. Depending on the application, different choices are possible. Here, we distinguish between momentum space domains considered as ``bulk'' and ``tail'', depending on the number of REs in each momentum bin. For each volume bin, our approach puts half of the markers into the ``bulk'' momentum space domain and the other half into the ``tail'' domain. The weights of the markers are adjusted accordingly. Figure~\ref{fig:avares_momentum}d) and g) show that the agreement of the resampled distribution with the initial one can be improved substantially this way. 
Compared to the naive resampling, the error in the total energy drops from $0.2\%$ to $0.0003\%$. For the same cases that were shown in figure~\ref{fig:avares_momentum}, the spatial distributions are analyzed in figure \ref{fig:nre_avares}. It can be seen that both resampling methods approximate the spatial distribution better than direct regeneration.

\section{JET-like scenario}
\label{sec:application}
With the implementation of the knock-on collisions and resampling as described in subsection~\ref{sec:moller} and subsection~\ref{sec:resampling}, the kinetic model in JOREK is now capable of not only evolving the REs in 3D fields, but also of simulating the secondary RE generation. 
For this purpose, a JET-like termination scenario was chosen as a first demonstration of the avalanche model. 
In previous work, it was shown that despite the stochastic fields, some REs remain confined in the core or magnetic islands and are still present at the end of the termination event when the flux surfaces reform~\cite{Bergstroem2025}.
Once confined, these REs could potentially re-avalanche if the electric field is strong enough, making this a useful case for a first application of the avalanche source. 

The JET-like termination scenario that will be considered here finds its original basis in shot \#95135 from the JET tokamak~\cite{Reux2021}. 
A beam of REs carrying a current of approximately 1 MA was generated using argon injections and was subsequently terminated benignly by injecting deuterium after the formation of the RE beam.
The neutral deuterium injection cools the background plasma further and reduces the ionization source term. 
This is accompanied by the recombination of argon ions, which greatly increases the radial transport of the argon out of the RE beam, as it is now dominated by neutral transport~\cite{Hollmann2020}. 
Eventually a violent MHD instability distributes the REs broadly across the plasma facing components without substantial hotspots.
This scenario had already been studied extensively in JOREK, first using a fluid RE model~\cite{Bandaru2021} and then with the kinetic-MHD hybrid model~\cite{Bergstroem2025}. 
In the experiment itself, there is little to no regeneration of REs, which is an important aspect to successfully achieve a benign termination of the discharge~\cite{Reux2021}. 
In this demonstration, we move away from the actual experiment and use the setup merely as a base for a JET-like termination, allowing for regeneration of REs while still building on the foundation laid by Bandaru \textit{et al.}~\cite{Bandaru2021} and Bergstr\"om \textit{et al}~\cite{Bergstroem2025}. 

The simulation is based on the profiles determined in Ref.~\cite{Bandaru2021}, to which the interested reader is referred for more details. 
In brief summary, the starting point of the simulation was chosen to be representative of the plasma state a few milliseconds before the termination.
At this time, most of the argon impurities were already lost in the experiment, motivating the choice of not modeling these impurities. 
Possible neutrals in the plasma are not considered either. 
The resistivity and viscosity are set to be spatially uniform, with: $\eta = \SI{3.3e-5}{\ohm\meter}$ (which corresponds to a Spitzer value at $T= \SI{10}{\electronvolt})$ and $\mu_\text{viscosity} = \SI{1.6e-6}{\kilogram\meter^{-1}\second^{-1}}$. 
The plasma current is assumed to be carried fully by the REs and is set to $I_p = I_\text{RE} = 0.747$ MA.
The synthetic diagnostic code SOFT~\cite{SOFT} had been used to constrain the current density and q-profile in such a way that the synchrotron radiation obtained from the distribution agrees with the experiment qualitatively, while making sure that the resulting MHD activity is dominated by a $(m,n) = (4,1)$ double tearing mode in line with experimental observations. 
As this work continues from the simulations in Ref.~\cite{Bergstroem2025}, the on-axis magnetic field is set higher than was used in Ref.~\cite{Bandaru2021} and is $B= \SI{3.601}{\tesla}$ which was chosen to compensate for a small change in q-profile induced by the switch to a kinetic representation for REs.

\begin{figure*}[p]
    \centering
    \includegraphics[width=.7\linewidth]{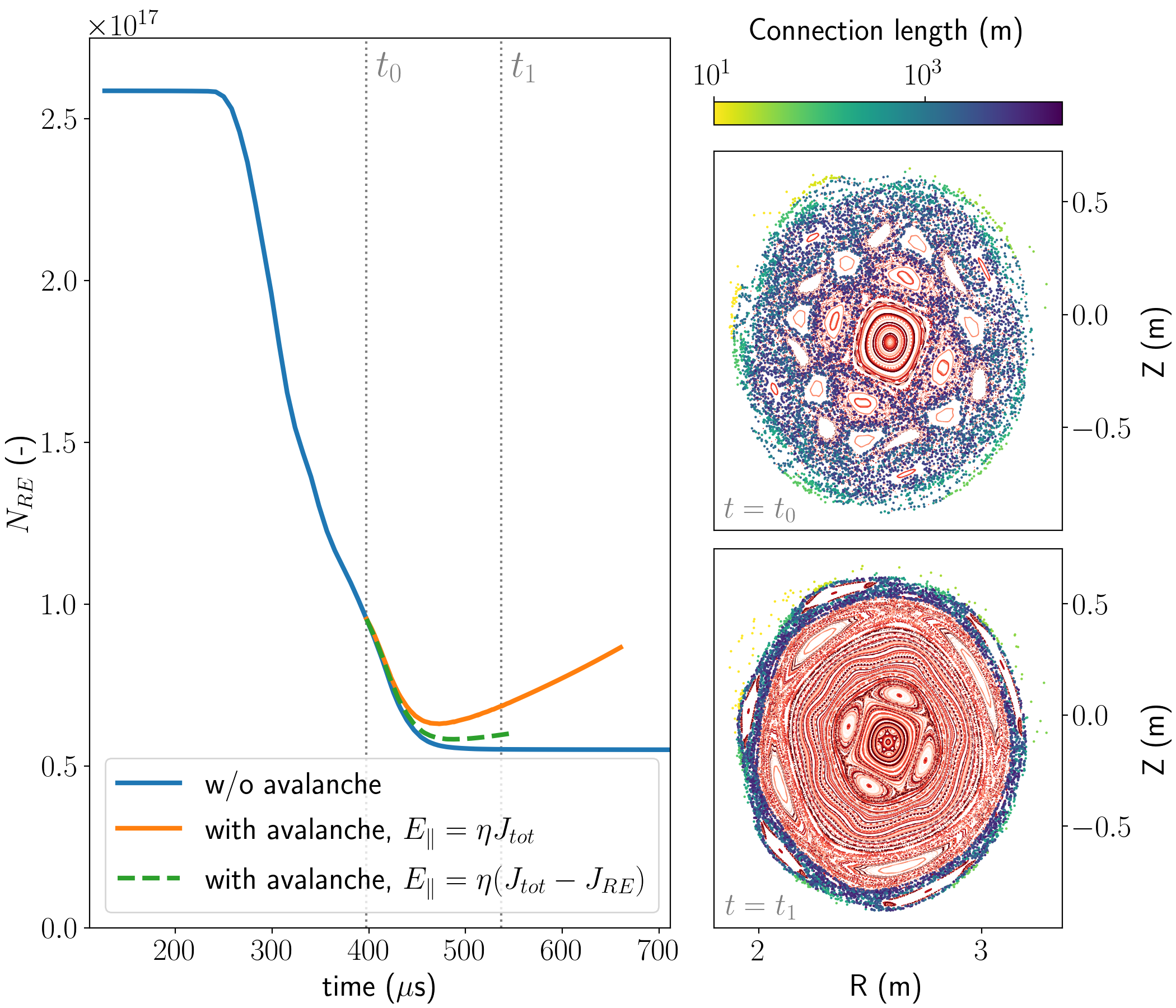}
    \caption{At time $t_0$ the simulation is restarted with the avalanche source included. The stochasticity in the field at this time as shown by the Poincar\'{e} plot is too high for the avalanche to overcome the RE losses. Towards the end of the simulation at $t_1$ the flux surfaces have reformed and the REs are confined again, allowing the avalanche source to dominate over the losses. When the simulated current is corrected for the current carried by the REs, $E_\parallel$ is lower, causing the avalanche gain to be lower as well.}
    \label{fig:reavalanching}
    \vspace{0.3cm}
    \centering
    \begin{tabular}{c@{\hspace{-4.3cm}}c}
        \begin{subfigure}[b]{0.45\textwidth}
            \centering
            \includegraphics[clip, trim=4cm 5cm 14cm 5cm, height=3.cm]{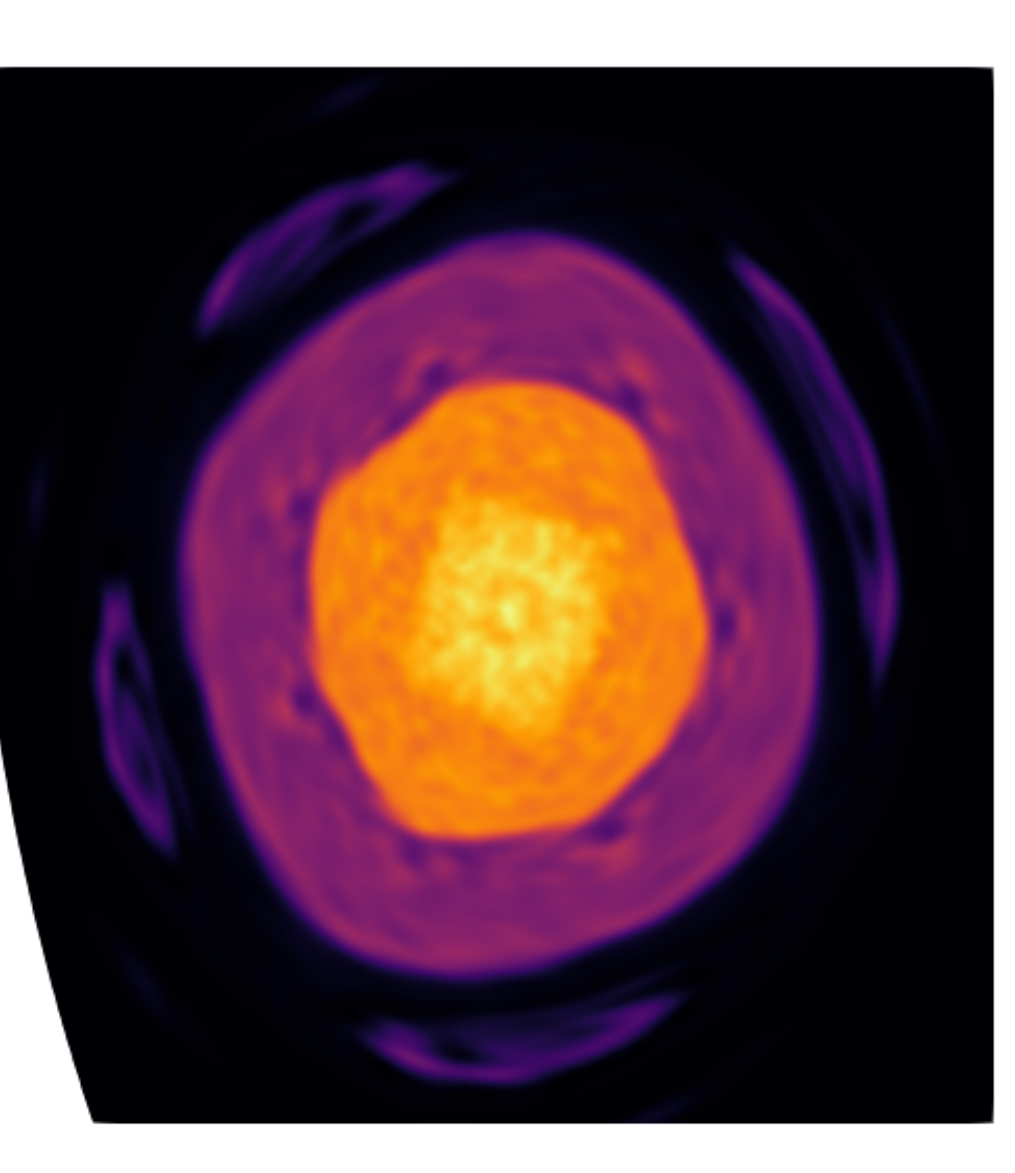}
            \caption{W/o avalanche}
        \end{subfigure}
        &
        \begin{subfigure}[b]{0.45\textwidth}
            \centering
            \includegraphics[clip, trim=4cm 5cm 3cm 5cm, height=3.cm]{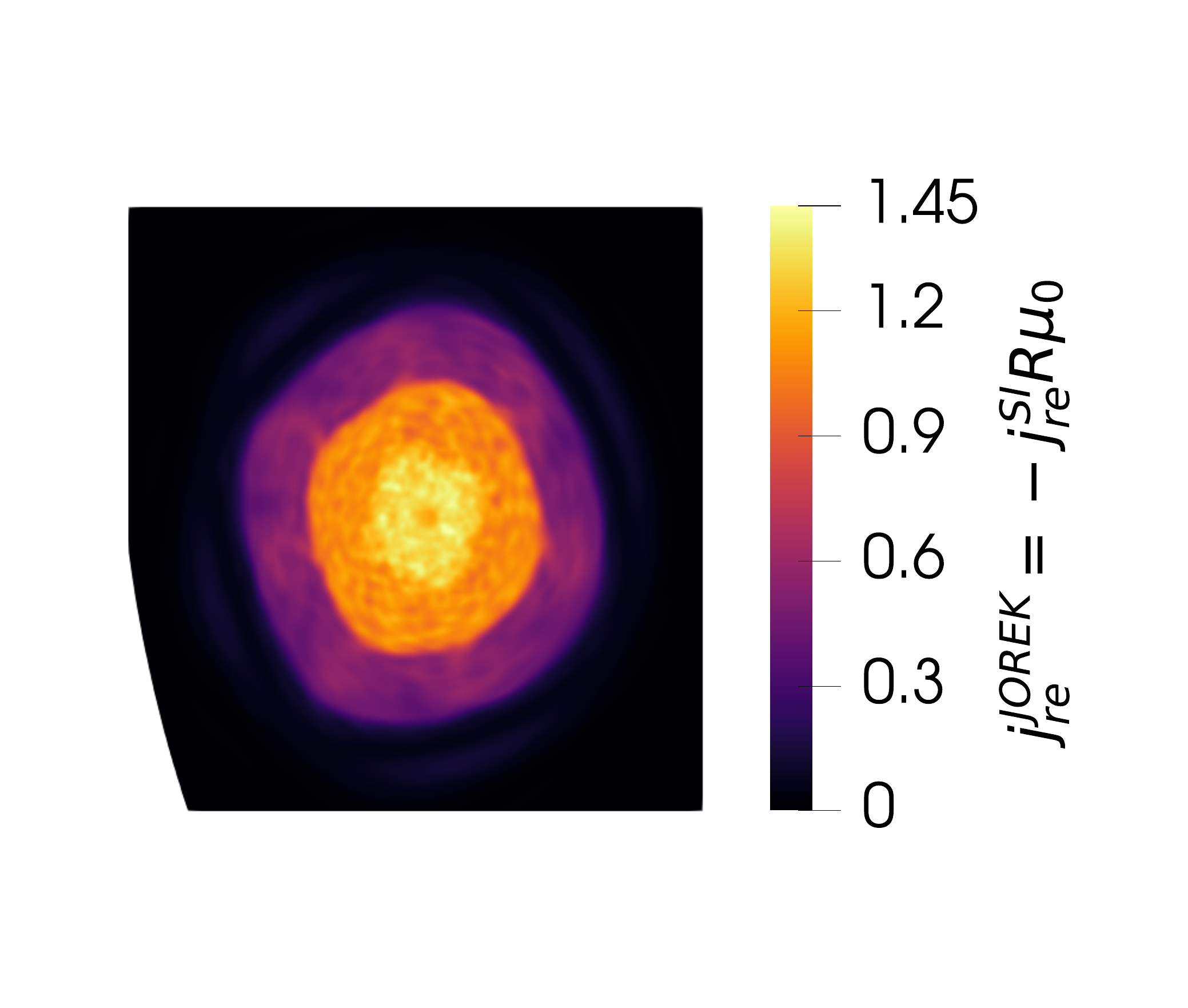}
            \caption{\text{With avalanche \hspace{1.7cm}}}
        \end{subfigure}
        \\
        &
        \begin{subfigure}[b]{0.45\textwidth}
            \centering
            \includegraphics[clip, trim=4cm 5cm 3cm 5cm, height=3.cm]{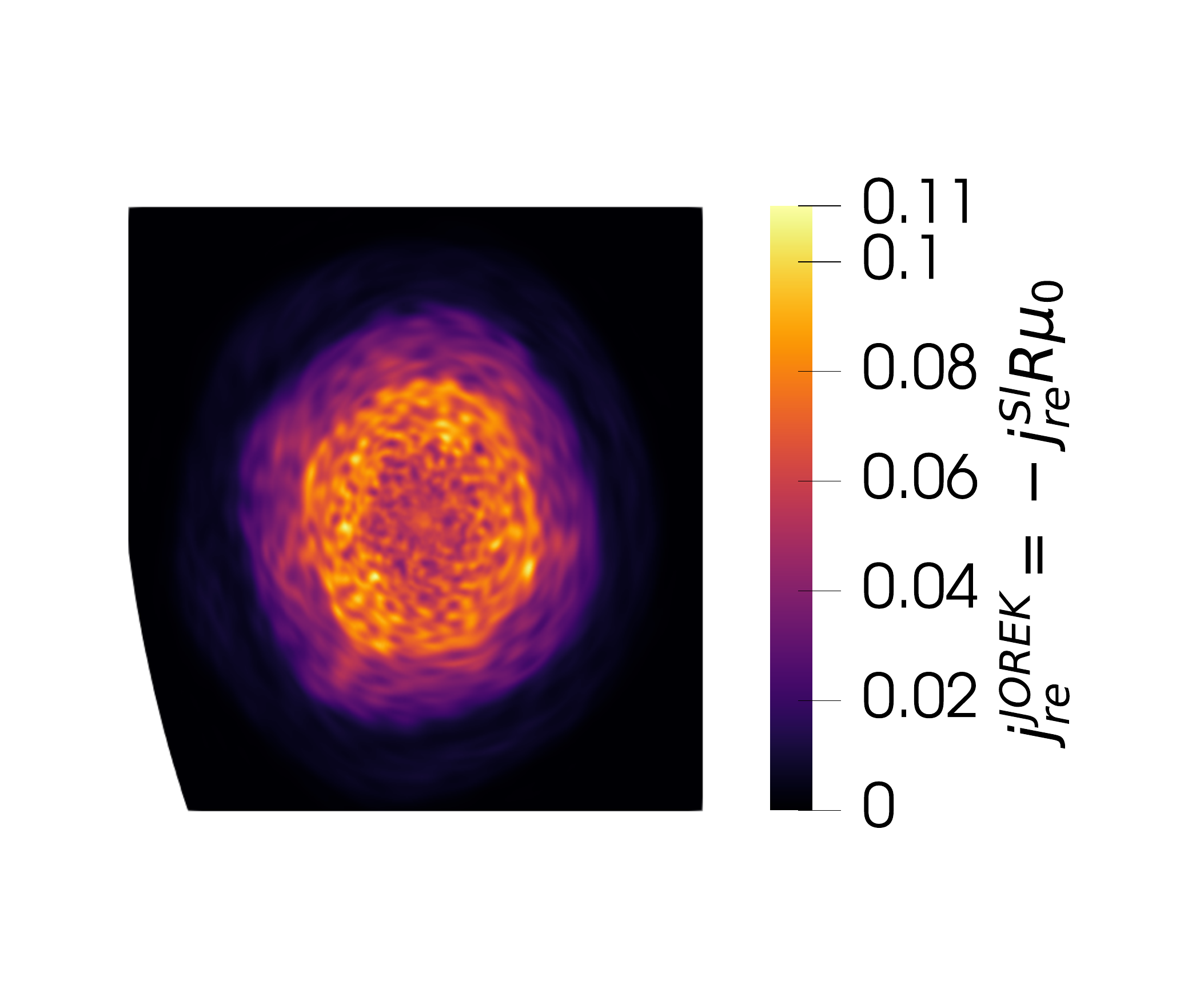}
            \caption{\text{Secondary $J_{RE}$ \hspace{1.6cm}}}
        \end{subfigure}
    \end{tabular}
    \nextfloat\caption{The current density carried by the REs at time $t = t_1$. The total current density for the cases without and with the avalanche (where $E_\parallel = \eta(J_\text{tot}-J_\text{RE})$) is shown in (a) and (b) respectively. In (c) the current density carried only by the secondary REs is shown. It can be seen that the current density of the REs is higher in a ring around the core, which corresponds to the region where the electric field is higher, corresponding to increased generation and acceleration of secondary electrons in this area.}
    \label{fig:j_RE_secon}
\end{figure*}

The initial marker population in this application is based on the simulations by Ref.~\cite{Bergstroem2025}, where $10^7$ markers with energy of 2MeV and pitch $\xi = 0.99$ were sampled proportional to the runaway density resulting from the fluid simulations in Ref.~\cite{Bandaru2021}.
As these simulations were also done with the JOREK code, the initial markers in this report can easily be set equal to those simulated by Ref.~\cite{Bergstroem2025} at any given initialization time. 
Since the purpose of including the avalanche is to simulate possible re-avalanching, the simulation is started towards the end of the termination (at $t=t_0$ in figure~\ref{fig:reavalanching}) when the flux-surfaces are close to reforming. 
Although the current is fully carried by REs before the onset of the termination, a significant fraction of REs is lost during the MHD burst so that the majority of the plasma current is converted back into Ohmic current, with REs just carrying a small fraction of the current by the time we start our simulation~\cite{Bergstroem2025}.
The fields and particles at this time are set equal to the simulations shown in Ref.~\cite{Bergstroem2025}, but from here on, the simulation will run one-way coupled, effectively tracing the REs in the electric and magnetic fields from the previously published coupled simulation, which is justified since the plasma current is primarily of Ohmic nature in this phase. No additional implementation is necessary for combining the hybrid fluid-kinetic model with our newly developed avalanche operator, but we decided against using it here for computational reasons (which we aim to reduce for future applications via numerical optimizations and GPU porting).
While in Ref.~\cite{Bergstroem2025} the REs were kept at a constant energy of 2 MeV, the energy and momentum of the REs will now be allowed to evolve by including the following effects: acceleration by the electric field, small-angle collisions, radiation reaction force and the effect of the knock-on collisions that make up the avalanche. 
At $t=t_1$ in figure~\ref{fig:reavalanching}, the simulation from Ref.~\cite{Bergstroem2025} stops and the fields in our demonstration are thus kept constant from that point onwards. 

It will be assumed that the free electron density is significantly lower than the bound electron density during the reformation phase of the simulation, as the recombination of the background plasma caused the free electron density in the experiment to drop below measurable values~\cite{Reux2021}. 
The free electron density is set to a constant value of $1\times10^{18} \text{ m}^{-3}$ and the bound electron density to a value of $1\times10^{19} \text{ m}^{-3}$ where this last value is chosen to match the background electron density used in Ref.~\cite{Bandaru2021}.
As the plasma is cold, temperature of the free electrons will also be assumed to be low, at a value of \SI{10}{\electronvolt}. 

Care was taken in choosing the cut-off parameter $\gamma_\text{min}$ for the large angle collisions, as the electric field varies both in space and time. 
Ideally, $\gamma_\text{min}$ is smaller than the local critical momentum $\gamma_c (R,Z,\phi) = \sqrt{1+\left(\frac{p_c(R,Z,\phi)}{m_e c}\right)^2}$, as the local electric field may vary before the secondary electron has significantly lost the energy gained in the large angle collision and fallen back into the bulk. 
Nevertheless, setting $\gamma_\text{min}$ too small will lead to the generation of a lot of markers that do not represent REs, as they will feel no net acceleration from the electric field. 
Such markers would increase computational cost and decrease the resolution for the RE population. 
For these simulations, $\gamma_\text{min}$ is set to a value smaller than the smallest value for $\gamma_c$ within the simulation domain and is kept constant over time. 
Due to the high electric field and low temperature of the plasma, this choice still ensures that $p_\text{min} = m_e c\sqrt{\gamma_\text{min}^2 -1} \gg p_\text{th}$. 
Nevertheless, this does not mean that double counting is completely avoided as the Coulomb logarithm should be modified to decrease the strength of the small-angle collisions as described in Ref.~\cite{Embreus2018}.
We leave this correction for future work, which will artificially increase the friction experienced by the tracer particles, but will still allow us to show re-avalanching with the new source.

As $\gamma_\text{min} < \gamma_c$ not all markers will represent REs, which are difficult to define precisely during the simulation due to the problem previously described. 
A way to solve this apparent dilemma is to take yet another energy cut-off $\gamma_\text{RE}$ and to consider all markers with energies above $\gamma_\text{RE} m_e c^2$ to represent REs. 
This third cut-off can be chosen such that $\gamma_\text{RE} > \gamma_c$ throughout the simulation in the region in space and time where $\gamma_c < \infty$. 

Switching from the coupled simulation of Ref.~\cite{Bergstroem2025} to particle tracing means that the current carried by the REs is no longer self-consistently included in the calculation of the field.
The parallel electric field is given by $E_\parallel = \eta J_\text{Ohm}$ with $\eta$ the resistivity and $J_\text{Ohm}$ the Ohmic current density in reality, but if the RE current from the kinetic simulation is not considered, the field is calculated instead as $E_\parallel = \eta (J_\text{Ohm} + J_\text{RE}) = \eta J_\text{tot}$. 
This artificially increases the electric field and so artificially increases the avalanche gain, causing a rapid re-avalanching once the flux-surfaces reform as seen in figure~\ref{fig:reavalanching}. 
A more correct approach is to subtract the RE current density from the total current density one would otherwise calculate, where we use $J_\text{RE}$ from the original coupled simulation in Ref.~\cite{Bergstroem2025}, which holds in the limit that the increase in $J_\text{RE}$ due to the RE avalanche is not significant. 
In this case $E_\parallel$ is lower and the avalanche growth is reduced, even though the start of re-avalanching can still be observed close to $t=t_1$.

On the relatively short timescales involved in this simulation, the secondary REs generated by the RE avalanche have not yet been able to gain enough energy to start overlapping with the now also accelerated and broadened initial population. 
The two populations are therefore clearly distinguishable, allowing one to investigate in which region the avalanche has the most impact on the current density. 
It is shown in figure~\ref{fig:j_RE_secon} that even though the total current density is highest in the core, the current density carried by the secondary REs is mainly localized in a ring around the core. 
This can be explained by the electric field being stronger outside of the core, which means both the growth rate as well as the energy gain of the secondary electrons over time were higher.

\section{Summary and conclusion}
\label{sec:conclusion}
In this work, a novel high-fidelity model for the runaway electron avalanche source has been implemented for the relativistic full-f particle-in-cell module of the 3D nonlinear MHD code JOREK. This allows to accurately capture RE generation and transport in the presence of 3D MHD fields, while fully resolving RE phase space dynamics, capturing the effects of trapped particles, and accounting for the self-consistent interaction between MHD modes and the kinetic RE orbits. In this article, we describe the implementation and verification of the avalanche source and the PiC resampling algorithm, along with a first relevant demonstration of the novel capabilities.

The avalanche itself is modeled with the use of a binary knock-on collision operator between a relativistic electron and an electron that is assumed to be approximately stationary.
These knock-on collisions are both energy and momentum conserving and allow for accurate modeling with respect to the possibility of collisions leading to particles with a high pitch-angle, which might be trapped or even traveling in the direction opposite to the electric field.
As it takes a long time for fast electrons to lose their energy, these particles are only removed once their energy has dropped below a set minimum, as they can still contribute to the avalanche until that time. 
The conservation of energy and momentum comes at the cost of a rapid, exponential increase in markers, which becomes computationally too expensive to run. 
A resampling technique has thus been implemented as part of the avalanche source, to complement the knock-on collisions. 

The implementation of the 5D resampling (3D + 2P) is based on the position of the guiding center in $(R,Z, \phi)$ coordinates and on the parallel momentum and magnetic moment $(p_\parallel, \mu)$ in momentum space. 
The spatial grid of the resampling is based on the JOREK grid used for the fluid simulation, with the added possibility of increasing the resolution and with additional equidistant toroidal bins, while the momentum grid is made up of equidistant bins spanning the momentum space of the group of markers within one volume bin. 
It was shown that the resampled distribution still exhibits the same 3D structures, even though the number of markers is significantly reduced and that, globally (not locally), the momentum binning conserves energy and momentum relatively well. 
Furthermore, the error scaling is as would be expected and the user has the freedom to choose between using the built-in pseudorandom and the quasi-random number generators. 

Using a JET-like termination scenario, we have shown that the combination of the knock-on collisions and the resampling can be used to model re-avalanching in an MHD-active scenario. 
Not taking into account that the electric field drops as the RE current increases leads to a greater avalanche gain.
Nevertheless, the case where the field was corrected for the RE current also shows an increase in REs and the start of re-avalanching. 
The secondary REs are generated at places where there is a higher RE density and a high parallel electric field and are able to gain more energy in these regions. 
As the simulation continues and most of the REs are confined in the core, the RE current density generated by the avalanche also starts increasing due to the high RE density there. 

Although not shown here to reduce computational costs, this model has been implemented in such a way that it can easily be used with self-consistent coupling to the MHD fields. 
This work therefore constitutes an important mile stone towards to a more complete model of the interaction of REs with the MHD during disruptions, which is applicable to scenarios that could previously only be studied using fluid models. Both fluid models and commonly used non-conservative avalanche sources have clear limitations in phase space transport of the REs that this model aims to overcome with a conservative knock-on collision operator for the kinetic model of the REs.
Applications to 3D MHD-active dynamics involving REs in high-current tokamaks like ITER are in preparation.
Further optimizations of the model as well as porting to accelerated high performance computing systems is also in preparation to allow crossing the long time scales that are relevant. 
Finally, nuclear sources will be implemented for kinetic RE studies in activated devices.

\section*{Data availability statement}
The data that support the findings of this study are available from the corresponding author upon reasonable request.

\section*{Acknowledgements}
This work has been carried out within the framework of the EUROfusion Consortium, funded by the European Union via the Euratom Research and Training Programme (Grant Agreement No 101052200 — EUROfusion). Views and opinions expressed are however those of the author(s) only and do not necessarily reflect those of the European Union or the European Commission. Neither the European Union nor the European Commission can be held responsible for them.
Some of the simulations were carried out on the Pitagora supercomputer hosted by Cineca and on the HPC systems TOK and RAVEN hosted by MPCDF. 
The authors are grateful to Dr P. Aleynikov for useful discussions.

\appendix
\counterwithin*{equation}{section}
\renewcommand\theequation{\thesection\arabic{equation}}
\section{Parallel and hyper filters}
\label{A:filters}
The implementation and use of the parallel and hyper filters used for smoothing the projections of the particle moments of the distribution function in JOREK have already been described in Refs.~\cite{Dvornova2021, Vugt2019}, but will be shortly summarized here for easy reference. 
The interested reader is always referred to Refs.~\cite{Dvornova2021, Vugt2019} for more details on the implementation and the effect of the smoothing. 

In order to map the moments of the distribution function onto the finite elements a projection $P$, which is a mapping such that $P^2 = P$, is used. 
The mapping is defined by writing an equation for the quantity to be projected $X$ and the projection $P$ in the weak form as follows,
\begin{equation}
\label{eq:genproj}
    \int (P\nu ) dV = \int X \nu dV,
\end{equation}
where $\nu$ is a test function. 
As the spatial position of the particles is described by a $\delta$-function, the integral on the right hand side of the equation reduces to to a sum. 
In order to reduce the noise in the projection, smoothing terms are added to the left hand side of equation \eqref{eq:genproj}.
\begin{align}
    \begin{split}
        \int(P \nu + \zeta\nabla P\cdot \nabla \nu + \zeta_\text{hyp} \nabla^2 P \nabla^2\nu + \zeta_\parallel\nabla_\parallel P \cdot \nabla_\parallel \nu)dV \\
        = \sum_i \delta(\mathbf{x} - \mathbf{x}_i) w_i X \nu.
    \end{split}
\end{align}
Here $\zeta, \zeta_\text{hyp}, \zeta_\parallel$ are the smoothing parameters, $\mathbf{x}_i$ is the position of particle $i$ and $w_i$ its weight. 
$P$ is defined as 
\begin{equation}
    P(\mathbf{x}) = \sum_{ijk} p_{ijk} H_{ij}(s,t)H_{\phi,k}(\phi),
\end{equation}
where $H_{ij} (s,t)$ are the finite element basis functions in the poloidal plane and $H_{\phi,k}$ are the toroidal Fourier components. 
The effect of the smoothing parameters is to diminish the amplitude of the higher mode numbers, therefore smoothing the projected values.

\bibliographystyle{IEEEtranNdoi}
\bibliography{references.bib}

\end{multicols}

\end{document}